\documentclass[%
aps,
prx,
reprint,
superscriptaddress,
nofootinbib,
amsmath,
amssymb
]{revtex4-2}
\usepackage{
    physics,
    graphicx,
    multirow,
    tabularx,
    mathtools, 
    placeins, 
    nicefrac, 
    hyperref, 
    threeparttable, 
    siunitx,
    enumitem,
    makecell
}
\usepackage{algorithm}
\usepackage{algorithmicx}
\usepackage{algpseudocode}
\usepackage[caption=false]{subfig}
\usepackage[capitalize]{cleveref}
\Crefname{section}{Sec.}{Secs.}

\usepackage[dvipsnames]{xcolor}
\hypersetup{
    colorlinks,
    linkcolor={blue!50!black},
    citecolor={blue!50!black},
    urlcolor={blue!80!black}
}

\usepackage[caption=false]{subfig} 
\usepackage{svg}
\usepackage[T1]{fontenc}
\DeclarePairedDelimiter\pbra{\langle\!\langle}{\rvert}
\DeclarePairedDelimiter\pket{\lvert}{\rangle\!\rangle}
\DeclarePairedDelimiterX\pbraket[2]{\langle\!\langle}{\rangle\!\rangle}{#1 \delimsize\vert #2}
\DeclarePairedDelimiterX\pketbra[2]{\lvert}{\rvert}{#1 \rangle\!\rangle\!\langle\!\langle #2}
\newcommand{\Binom}[2]{\ensuremath{\binom{#1}{#2}}}

\makeatletter
\AddToHook{cmd/appendix/before}{\def\cref@section@alias{appendix}\def\cref@subsection@alias{appendix}}
\makeatother

\begin{document}

\title{Quantum Error Correction on Error-mitigated Physical Qubits}

\newcommand{\qmaddress}{\affiliation{Quantum Motion, 9 Sterling Way, London N7 9HJ, United Kingdom}}
\newcommand{\oxddress}{\affiliation{Department of Materials, University of Oxford, Parks Road, Oxford OX1 3PH, United Kingdom}}

\newcommand{\oxengaddress}{\affiliation{Department of Engineering Science, University of Oxford, Parks Road, Oxford OX1 3PJ, United Kingdom}}

\author{Minjun Jeon}
\email{minjun.jeon@materials.ox.ac.uk}
\oxddress
\qmaddress

\author{Zhenyu Cai}
\email{cai.zhenyu.physics@gmail.com}
\oxengaddress
\qmaddress

\date{\today}

\begin{abstract}  
    We present a general framework for applying linear quantum error mitigation (QEM) techniques directly to physical qubits within a logical qubit to suppress logical errors. By exploiting the linearity of quantum error correction (QEC), we demonstrate that any linear QEM method---including probabilistic error cancellation (PEC), zero-noise extrapolation (ZNE), and symmetry verification---can be integrated into the physical layer without requiring modifications to the subsequent QEC decoder. Applying this framework to memory experiments using PEC, we analytically prove and numerically verify that the leading-order contribution to the logical error can be removed, increasing the effective code distance by 2. Our simulations on repetition and rotated surface codes show that a distance-3 code with physical-level PEC achieves logical error rates lower than or similar to a distance-5 unmitigated code while using 40\% and 64\% fewer qubits, respectively. These results establish physical-level QEM as a widely compatible and resource-efficient strategy for enhancing logical performance in early fault-tolerant architectures.
\end{abstract}

\maketitle

\section{Introduction}
Reducing the impact of noise is the central challenge on the path from proof-of-principle demonstrations to practical quantum computation.
Two complementary approaches have recently emerged.
\emph{Quantum error correction} (QEC) protects information by encoding it into many physical qubits and decoding syndromes to suppress errors~\cite{gottesman1997stabilizer,Fowler_2012,Terhal_2015}, while \emph{quantum error mitigation} (QEM) reduces bias in measured observables by combining multiple noisy circuit executions without substantial qubit overhead~\cite{Temme_2017,Li_2017,Cai_2023}.
Recent experiments have made rapid progress on both fronts, with surface-code demonstrations that improve logical performance as the distance increases~\cite{krinner2022realizing,google2023suppressing,google2025quantum,Campagne_Ibarcq_2020} and with error-mitigation techniques operating at scale~\cite{kandala2019error,kim2023scalable,van_den_Berg_2023}. Together, these developments signal the onset of an \emph{early fault-tolerant} era in which devices host tens of logical qubits and support increasingly complex protected circuits.

In the near term, with limited physical-qubit budgets, modest code distances, and finite measurement rounds, neither approach alone is sufficient for many tasks~\cite{piveteau2021error, Takagi_2022, mohammadipour2025directanalysiszeronoiseextrapolation}. Therefore, for the foreseeable future, a hybrid strategy that combines quantum error correction (QEC) with quantum error mitigation (QEM) is essential to fully exploit the computational power of any quantum machines we have. The first set of explorations in this direction applied QEM at the logical level, targeting logical errors through modifying and sampling from logical circuits~\cite{piveteau2021error,suzuki2022}. It is especially useful in targeting bottlenecks in logical computation like magic state factories and compilation errors. Further work has seen the development of virtual error correction, which is the first example of native integration of QEC and QEM~\cite{liu_2025,araki2025correctingquantumerrorsusing} that are not simply a concatenation of one on top of another, allowing the correction of quantum errors using classical codes and only one additional qubit. 

To apply QEM at the logical layer, all of the device calibration and modification of quantum circuits need to happen at the logical level, which can be significantly slower and more challenging than the corresponding physical operations. Therefore, it is natural to ask whether we can apply QEM directly on the physical qubits and then QEC on top of the error-mitigated physical layer. This direction has been far less explored due to the perceived complexity of the interaction between modification in the physical circuit and the decoding process of QEC on top. It was realised in Ref.~\cite{liu_2025} that no decoder modification is needed when applying probabilistic error cancellation (PEC) at the physical level in the context of virtual error correction. Zhang et al.~\cite{zhang2025demonstratingquantumerrormitigation} also experimentally demonstrated that applying zero-noise extrapolation (ZNE) by amplifying noise on the physical qubits can reduce logical errors. 

In this paper, we have constructed a general framework showing that any linear QEM methods, which include most mainstream QEM techniques such as PEC~\cite{Temme_2017,Endo_2018}, Richardson ZNE~\cite{Li_2017,Temme_2017}, symmetry verification~\cite{mcardleErrormitigatedDigitalQuantum2019,bonet-monroigLowcostErrorMitigation2018} and virtual purification~\cite{Koczor_2021,Huggins_2021}, can be implemented at the physical level to reduce the logical error rate of any QEC process applied on top. We then study the explicit example of using PEC to cancel out the leading-order physical errors in the repetition code and surface code, both analytically proving and numerically demonstrating an increase in the effective code distance for both codes. 

In \cref{sec: two_equiv_pictures}, we will begin by constructing a framework that rigorously shows any linear QEM method can be applied to physical qubits within any QEC code. We will then apply this framework to PEC in a memory experiment in \cref{sec: memory} and derive a general expression for the error-mitigated logical error rate. In \cref{sec: numerics}, we present numerical simulations for repetition and rotated surface codes, compare them with theoretical predictions, and discuss threshold behaviour. \cref{sec: conclusions} summarises the results and outlines future directions.

\section{QEC on Error-Mitigated Physical Qubits} \label{sec: two_equiv_pictures}
\subsection{Linear QEM: definition and examples}\label{subsec:linear-qem}
As discussed in \cite{caiPracticalFrameworkQuantum2021}, a QEM method is called linear if the expectation value of the error-mitigated estimator $\hat{O}_{\mathrm{em}}$ is linear in the observable of interest $O$ and thus can be written in the form of 
\begin{align*}
         \mathbb{E}[\hat{O}_{\mathrm{em}}] &= \Tr(O\rho_{\mathrm{em}})
\end{align*}
for some error-mitigated ``state'' $\rho_{\mathrm{em}}$. This effective ``state'' might not satisfy the definition of a physical quantum state (it is usually unit trace, but might not be positive).

For practical implementation, the error-mitigated expectation value is constructed from the linear combination of the output from a set of circuit configurations labelled using index $b$:
\begin{align}\label{eqn:qem_implement_1}
    \mathbb{E}[\hat{O}_{\mathrm{em}}] = \Tr(O \rho_{\mathrm{em}}) = \sum_{b} \beta_b \Tr((O \otimes I) S_b \rho_b).
\end{align}
Here for the given circuit configuration $b$, $\rho_b$ is the output state which may contain additional auxiliary qubits on top of the main register that $O$ acts on. $(O \otimes I) S_b$ is the measured observable with $I$ acting on the auxiliary system and $\beta_b$ is the corresponding weight assigned to the circuit configuration. The exact circuit is shown in \cref{fig:circ_implementation}. The formalism here covers all mainstream QEM techniques such as PEC, Richardson ZNE, symmetry verification and virtual purification. \cref{eqn:qem_implement_1} implies that the error-mitigated state is given by:
\begin{align*}
    \rho_{\mathrm{em}} = \sum_{b} \beta_b \Tr_{\mathrm{aux}}(S_b \rho_b)
\end{align*}
where $\Tr_{\mathrm{aux}}$ is the partial trace over the auxiliary registers that are not acted on by $O$.

For ease of derivations later, we will rewrite \cref{eqn:qem_implement_1} using the process matrix notation, where operators like $\rho_{\mathrm{em}}$ are written as vectors like $\pket{\rho_{\mathrm{em}}}$:
\begin{align}\label{eqn:qem_implement}
    \mathbb{E}[\hat{O}_{\mathrm{em}}] = \pbraket{O}{\rho_{\mathrm{em}}} = \sum_{b} \beta_b \pbraket{O \otimes I}{S_b \rho_b}.
\end{align}

\subsection{Examples: Probabilistic Error Cancellation}\label{sec:pec_qem}
For a given noise channel $\mathcal{E}$, its inverse can be decomposed into the linear combination of physically implementable basis operations $\{\mathcal{B}_{b}\}$:
\begin{align*}
    \mathcal{E}^{-1} = \sum_{b} \alpha_b \mathcal{B}_{b}. 
\end{align*} 
Given the noisy state $\pket{\rho} = \mathcal{E}\pket{\rho_0}$, which is the ideal incoming state $\rho_0$ that gets corrupted by $\mathcal{E}$, we can recover the ideal expectation value of $\rho_0$ with respect to the observable of interest $O$ using:
\begin{align*}
    \pbraket{O}{\rho_0} &= \pbra{O}\mathcal{E}^{-1}\mathcal{E}\pket{\rho_0} = \sum_b \alpha_b \pbra{O}\mathcal{B}_{b}\pket{\rho}.
\end{align*}

More generally, if we are not targeting the perfect expectation value $\pbraket{O}{\rho_0}$, but simply some error-mitigated one $\pbraket{O}{\rho_\mathrm{em}}$, we can adjust the weights $\alpha_b$ to $\beta_b$ so that we can construct an approximate inversion process:
\begin{align}
    \mathcal{F} = \sum_{b} \beta_b \mathcal{B}_{b},
\end{align}
which only targets specific error components~\cite{caiMultiexponentialErrorExtrapolation2021}. The resultant error-mitigated expectation value is then

\begin{align}\label{eqn:pec_qem}
    \pbraket{O}{\rho_\mathrm{em}} &= \pbra{O} \mathcal{F} \circ \mathcal{E} \pket{\rho_{0}} \nonumber\\
    &= \pbra{O}\mathcal{F}\pket{\rho} = \sum_b \beta_b \pbra{O}\mathcal{B}_{b}\pket{\rho},
\end{align}where the composite channel, $\mathcal{F} \circ \mathcal{E}$, can be thought of as an \textit{effective error channel}. Compared to \cref{eqn:qem_implement}, we see that this corresponds to $\pket{\rho_b} = \mathcal{B}_{b}\pket{\rho}$, $S_b = I$.

The arguments above can also be generalised to the cases in which we insert gates in the middle of the noisy state preparation circuit to combat mid-circuit noise rather than only inserting gates at the end~\cite{Temme_2017,Endo_2018}. 

Similar arguments can also be extended to other QEM techniques. For ZNE~\cite{Li_2017,Temme_2017}, we have $S_b = I$ and  $\rho_b$ are states with different noise levels. For symmetry verification~\cite{mcardleErrormitigatedDigitalQuantum2019,bonet-monroigLowcostErrorMitigation2018}, $S_b$ is the symmetry projector (or the symmetry operators) and $\rho_b$ is simply the noisy state itself. For virtual purification~\cite{Huggins_2021,Koczor_2021}, $\rho_b$ is multiple copies of the noisy states and $S_b$ is the cyclic permutation operator among these copies.

\subsection{The QEC Channel} \label{subsec: linearity_QEC}
In QEC, given an incoming corrupted encoded state, we will first perform stabiliser checks to obtain the error syndrome $s$. This will project the incoming code state into the $s$-syndrome subspace with the corresponding projector channel denoted as $\mathcal{M}_{s}$. We can then apply the corresponding correction operation $\mathcal{C}_{s}$ output from the decoder to recover the resultant state back into the code space. The whole QEC process $\mathcal{R}$ is thus given as:
\begin{align}\label{eqn: qec_recovery_channel_def}
    \mathcal{R} = \sum_{s} \mathcal{C}_{s} \mathcal{M}_{s}
\end{align}

Using the process matrix notation, applying QEC on an incoming corrupted encoded state $\rho$ and then measuring the logical observable $O$ will give us the following error-corrected expectation value:
\begin{align}\label{eqn:qec_implement}
    \mathbb{E}[\hat{O}_{\mathrm{ec}}] = \pbra{O}\mathcal{R} \pket{\rho} = \sum_{s} \pbra{O}\mathcal{C}_{s} \mathcal{M}_{s} \pket{\rho}
\end{align}
where $\hat{O}_{\mathrm{ec}}$ is the estimator for the error-corrected expectation value.

\subsection{QEC on top of Error-mitigated Physical Qubits} \label{subsec: two_equiv_pictures_qem_qec}

\begin{figure}
    \centering
    \includegraphics[width=0.75\linewidth]{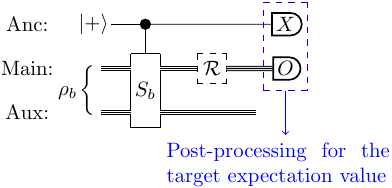}
    \caption{Circuit implementation of QEM at the physical level without QEC (without the dashed-boxed $\mathcal{R}$ channel) and with QEC on top (with the dashed-boxed $\mathcal{R}$ channel).}
    \label{fig:circ_implementation}
\end{figure}
If the noise on the encoded state is very strong or it is of some adversarial form, there can still be significant residual noise in the error-corrected state. In such a case, we may consider applying additional error mitigation procedures on top of QEC. As mentioned before, there have been attempts to apply QEM at the logical level~\cite{suzuki2022,piveteau2021error}, however it comes with associated challenges of logical noise characterisation, etc. 

Here we will study the possibility of performing QEM at the physical level of an encoded state. In this case it simply means that, before performing QEC, we will first perform error-mitigation directly on the incoming corrupted encoded state $\rho$, which is corrupted by noise on physical qubits and can be outside the code space. This turns the incoming state into an error-mitigated state $\rho_{\mathrm{em}}$, and using \cref{eqn:qec_implement} we can obtain the error-corrected-and-mitigated expectation value as 
\begin{align}\label{eqn: composite_channel}
    \mathbb{E}[\hat{O}_{\mathrm{emc}}] = \pbra{O}\mathcal{R} \pket{\rho_{\mathrm{em}}}
\end{align}
where $\hat{O}_{\mathrm{emc}}$ is the estimator for the error-mitigated-and-corrected expectation value. As long as $\rho_{\mathrm{em}}$ contains less noise that can escape the QEC recovery map $\mathcal{R}$ than the incoming corrupted code state $\rho$, then we can be sure that $\mathbb{E}[\hat{O}_{\mathrm{emc}}]$ will be more accurate than $\mathbb{E}[\hat{O}_{\mathrm{ec}}]$, i.e. the corresponding estimator $\hat{O}_{\mathrm{emc}}$ is less biased than $\hat{O}_{\mathrm{ec}}$. 

A natural concern is how such a QEM procedure can be implemented before the QEC process. It seems that all of the circuit modifications needed at the physical level will interfere with the error correction process and non-trivial adaptation to the decoders is needed. Here we will show that no modification to the QEC process is needed, i.e. the implementation of QEC is agnostic to the type of (or the absence of) QEM at the physical level. 

We will absorb the QEC process into the observable to define the new effective observable $\pbra{\widetilde{O}} = \pbra{O}\mathcal{R}$. Now using \cref{eqn:qem_implement} with $O \Rightarrow \widetilde{O}$, \cref{eqn: composite_channel} will become:
\begin{align}
    \mathbb{E}[\hat{O}_{\mathrm{emc}}] &= \pbraket{\widetilde{O}}{\rho_{\mathrm{em}}} = \sum_{b} \beta_b \pbraket{\widetilde{O} \otimes I}{S_b \rho_b} \nonumber\\
&= \sum_{b} \beta_b \pbra{O \otimes I}(\mathcal{R} \otimes \mathcal{I})\pket{S_b \rho_b} \label{eqn: branch_conditioned}
\end{align}

In practice, the expectation value in \cref{eqn: branch_conditioned} is evaluated by Monte Carlo sampling over the circuit configurations indexed by $b$. In each shot, a single configuration $b$ is drawn according to the probability $\abs{\beta_b}/\sum_{b}\abs{\beta_b}$, the corresponding circuit producing $\rho_{b}$ is executed, the QEC recovery channel $\mathcal{R}$ is applied based on the measured syndrome, and the logical observable $O$ is measured. The appropriate sign $\beta_b/\abs{\beta_b}$ is then attached to the result to give one sample of our error-mitigated estimator (see \cref{appendix: sec: operational_definition} for more details).

The implementation for each circuit configuration $b$ is shown in \cref{fig:circ_implementation}. There we see that we treat the incoming noisy state $\rho$ in the same usual way as in regular QEM, which will enable us to construct a series of circuit configurations indexed using $b$ for the purpose of QEM. No modification to the elements $\rho_b$, $S_b$ and $\beta_b$ of these circuit configurations is needed to account for the fact that we are applying on top of a noisy encoded state. The only modification we need to make is inserting the QEC process $\mathcal{R}$ before the measurement of the observable $O$. The QEC process $\mathcal{R}$ here is also not modified at all compared to the one for the unmitigated state $\rho$. 

Returning to the PEC example in \cref{sec:pec_qem}, applying QEC on top will turn \cref{eqn:pec_qem} into
\begin{align}\label{eqn:pec_qem_qec}
    \mathbb{E}[\hat{O}_{\mathrm{emc}}] &= \pbra{O}\mathcal{R}\mathcal{F}\pket{\rho} = \sum_b \beta_b \pbra{O}\mathcal{R}\mathcal{B}_{b}\pket{\rho}
\end{align}
We see that no modification to the error correction process $\mathcal{R}$ is required, even though we are inserting different gates into the circuit by running different $\mathcal{B}_{b}$. The unawareness of the decoder about the gate insert is precisely what makes PEC work: it introduces deliberate logical errors in individual circuit runs to achieve error cancellation when combining the results from different runs.

So far we are focusing on a single round of perfect QEC channels at the end, which is essentially the code-capacity model. In practice, the QEC channel itself is noisy, therefore we need to implement multiple rounds of such QEC channels.  In the discussion above we use the QEM framework in \cref{subsec:linear-qem} which covers most of QEM methods. As mentioned in Ref.~\cite{Cai_2023}, some QEM techniques can be viewed as transforming the noisy channel $\mathcal{U}_{\epsilon}$ into an effective error-mitigated process $\mathcal{U}_{\mathrm{em}}$ that is less noisy. Hence, if the QEC channel $\mathcal{R}_{\epsilon}$ is noisy, we will be able to construct an error-mitigated QEC map $\mathcal{R}_{\mathrm{em}}$ to reduce the noise inside. This error-mitigated map is linear: for PEC and ZNE (with Richardson extrapolation), it takes the form of a linear combination of noisy circuits, while for symmetry verification and virtual channel purification it is a quantum comb acting on the incoming noisy channel~\cite{liu_2025} (more generally it can be a virtual comb~\cite{zhuReversingUnknownQuantum2024}). Crucially, as a linear quantum map, $\mathcal{R}_{\mathrm{em}}$ remains composable. Therefore, this subclass of QEM strategy is directly applicable to consecutive multiple rounds of noisy QEC by concatenating the effective error-mitigated processes: $\mathcal{R}_{\mathrm{em}}^{(L)} \circ \cdots \circ \mathcal{R}_{\mathrm{em}}^{(1)}$. 

\section{PEC Framework for Memory Experiments} \label{sec: memory}
\subsection{Noise Inversion Process}\label{subsec: inversion_channel}
In this section, we will consider the application of PEC on the physical qubits of a QEC code. Assuming a code-capacity model with each qubit corrupted with the error probability $p$, the leading order logical errors of a QEC code with distance $d$ are caused by physical errors of weight-$\omega=\lceil d/2\rceil$, occurring with the probability $\order{p^{\omega}}$. As discussed in \cref{sec:pec_qem}, we can construct an approximate inversion process $\mathcal{F}$ using PEC to remove specific errors. Thus in this case, it is natural to try to target the weight-$\omega$ errors before implementing QEC, which will reduce the leading order error probability from $\order{p^{\omega}}$ to $\order{p^{\omega + 1}}$. 

Let us consider the case where each qubit suffers from some error channel $\mathcal{G}$ with probability $p$, which gives the overall error channel:
\begin{align} 
    \mathcal{E} &= \Bigr[\bigl (1-p \bigr )\mathcal{I} + p\mathcal{G}\Bigl]^{\otimes N} = \sum^{N}_{k=0} P_k \sum_{|K|=k} \mathcal{G}_{K}.\label{eqn: pauli_error_channel_weight_w}
\end{align}
where $K \subseteq[N]$ is a support set, denoting the locations of the errors, with $|K|$ being the weight of the error and
\begin{align}
    P_k = p^{k}(1-p)^{N-k}
    \label{eqn: weight_k_error_prob}
\end{align}
is the probability of a given weight-$k$ error happening. For example, we have $\mathcal{G} = \mathcal{X}$ for bit-flip noise and $\mathcal{G} = (\mathcal{X} + \mathcal{Y} + \mathcal{Z})/3$ for depolarising noise. 

A possible inversion process we can implement is:
\begin{equation} \label{eqn: approx_inv_channel}
    \mathcal{F} = \frac{1}{A}\left(P_0\mathcal{I} -P_\omega\sum\nolimits_{|K| = \omega}\mathcal{G}_{K}\right),
\end{equation}
with the normalisation factor $A = P_0 - \Binom{N}{\omega}P_\omega$, and the sum runs over all supports of weight-$\omega$ errors. Using PEC, the inverse channel is implemented by sampling the different circuit configurations with the appropriate sign attached, which will also be referred to as the different \emph{branches} in the PEC implementation. There is the identity branch which leaves the circuit unchanged, and the branches where additional noise process $\mathcal{G}_K$ are inserted with signed weights. The collection of all non-identity branches will also be called a \emph{``superbranch''} for the ease of reference later.

Noticing $P_k \sim \order{p^k}$, which means
\begin{align*}
   \left(P_\omega\sum\nolimits_{|K| = \omega}\mathcal{G}_{K}\right)\mathcal{E} = P_0 P_\omega \sum\nolimits_{|K| = \omega}\mathcal{G}_{K} + \order{p^{\omega+1}}
\end{align*}
Therefore, the effective error channel becomes
\begin{align}
   \mathcal{F} \circ \mathcal{E} &= \frac{P_0}{A}\left[\sum^{N}_{k=0} P_k \sum_{|K|=k} \mathcal{G}_{K} - P_\omega \sum_{|K| = \omega}\mathcal{G}_{K} + \order{p^{\omega+1}}\right] \nonumber\\
   & = \frac{P_0}{A} \sum^{N}_{k=0, k \neq \omega} P_k \sum_{|K|=k} \mathcal{G}_{K} + \order{p^{\omega+1}} \nonumber\\
   & = \frac{P_0}{A} \sum^{\omega - 1}_{k=0} P_k \sum_{|K|=k} \mathcal{G}_{K} + \order{p^{\omega+1}}\label{eqn: effective_err_channel_memory}\\
\end{align}
which indeed suppresses the probability of all errors with weight-$\omega$ and above from $\order{p^{\omega}}$ to $\order{p^{\omega+1}}$. The closed-form expression for the composite channel $\mathcal{F} \circ \mathcal{E}$ for the bit-flip channel, i.e. $\mathcal{G}=X$, and depolarising channel, i.e. $\mathcal{G} = \frac{1}{3}\bigl(\mathcal{X} + \mathcal{Y} + \mathcal{Z}\bigr)$, are derived and discussed in \cref{appendix: sec: derivation_err_mit_channel}.

Note that for the approximation above to be valid, we need to operate in a regime where higher-order errors are less likely, which implies that $A$ in \cref{eqn: approx_inv_channel} needs to be positive. $A$ will become zero at the physical error rate of
\begin{align}\label{eqn: pole}
    p_{\mathrm{pole}}=\frac{1}{\,1+\binom{N}{\omega}^{1/\omega}\,}
\end{align}
and thus we need to operate at a physical error rate $p < p_{\mathrm{pole}}$ to achieve noise suppression using the inversion channel above. \cref{sec:pole} shows more detailed analysis of $p_{pole}$ for repetition and surface code.

Such a noise inversion can be generalised into a circuit-level noise model by considering $K$ as the set of error locations spanning over different time steps rather than just at one time step, and the exact same arguments still apply. In this case, $\mathcal{E}$ will not be an error channel, but a more general quantum comb that is supported over the various error locations in the circuit. If error locations have physical error rates of different orders of magnitude, the notion of a single dominant weight $\omega$ becomes ambiguous, and the construction of an optimal inverse channel requires refinement.

\subsection{Logical Error Rate}
\label{subsec: log_err_rate}
Consider a memory experiment starting in the logical $\ket{0_{L}}$ state; the corresponding error-mitigated logical error rate can be obtained using \cref{eqn:pec_qem_qec} with the incoming state $\pket{\rho} = \mathcal{E}\pket{0_L}$ and the target observable $\pbra{O} = \pbra{1_L}$ being the projector into the logical $\ket{1_{L}}$ state:
\begin{align}\label{eqn:pec_qem_qec_2}
    P^{\mathrm{PEC}}_L(p;d) &= \pbra{1_L}\mathcal{R}\mathcal{F}\mathcal{E}\pket{0_L}
\end{align}

Let us now see more explicitly how the logical error rate is reduced by first considering the performance of the individual circuit configuration we run, and then considering how they are combined. This can provide more intuition of why no modification to the QEC process (and specifically the decoder) is needed. We will call this the branch-conditioned picture.

In the approximate noise inversion process in \cref{eqn: approx_inv_channel}, there are basically two components: an identity component and another component that introduces weight-$\omega$ errors, namely the superbranch component. Similarly, we can also split our error-mitigated logical error rate in \cref{eqn:pec_qem_qec_2} into these two components:
\begin{align}\label{eqn: log_err_rate_branch_cond_no_gain}
    P^{\mathrm{PEC}}_L(p;d) &= \frac{P_0}{A} P^{(0)}_{L} -  \binom{N}{\omega}\frac{P_\omega}{A}P^{(\omega)}_{L}
\end{align}
where 
\begin{align*}
    P^{(0)}_{L} = \pbra{1_L}\mathcal{R}\mathcal{E}\pket{0_L}
\end{align*}
is the unmitigated logical error rate and
\begin{align*}
     P^{(\omega)}_{L} = \pbra{1_L}\mathcal{R}\left(\sum\nolimits_{|K| = \omega}\mathcal{G}_{K}\right)\mathcal{E}\pket{0_L} \bigg/ \binom{N}{\omega}
\end{align*}
is the average logical error rate by deliberately introducing weight-$\omega$ physical errors into this circuit. Note that we do not need to adjust the decoders in order to improve the performance when we deliberately introduce weight-$\omega$ errors. We want the raw performance so that this can exactly cancel out the damage done by the same errors in the unmitigated branch. 

Using $A = P_0 - \Binom{N}{\omega}P_\omega$, we can rewrite \cref{eqn: log_err_rate_branch_cond_no_gain} into:
\begin{align}\label{eqn: log_err_rate_branch_cond}
    P^{\mathrm{PEC}}_L(p;d) &=  P^{(0)}_{L} -  \binom{N}{\omega} \frac{P_\omega}{A}(P^{(\omega)}_{L} - P^{(0)}_{L})
\end{align}
where $\binom{N}{\omega} \frac{P_\omega}{A}(P^{(\omega)}_{L} - P^{(0)}_{L})$ is the amount of reduction in the logical error rate by applying PEC.

In \cref{appendix: sec: derivation_err_mit_channel}, we explicitly show the cancellation of $\order{p^{\omega}}$ for the bit-flip noise and depolarising noise. The relationship between \cref{eqn: log_err_rate_branch_cond_no_gain} and the Monte Carlo estimator in \cref{eqn: log_err_rate_operational} is further explored in \cref{appendix: sec: operational_definition}.

\section{Numerics and Discussion}
\label{sec: numerics}

\begin{figure}[htbp]
    \centering
    \subfloat[\label{fig: stim_circuit_rc_vanilla}]{\includegraphics[width=0.5\textwidth]{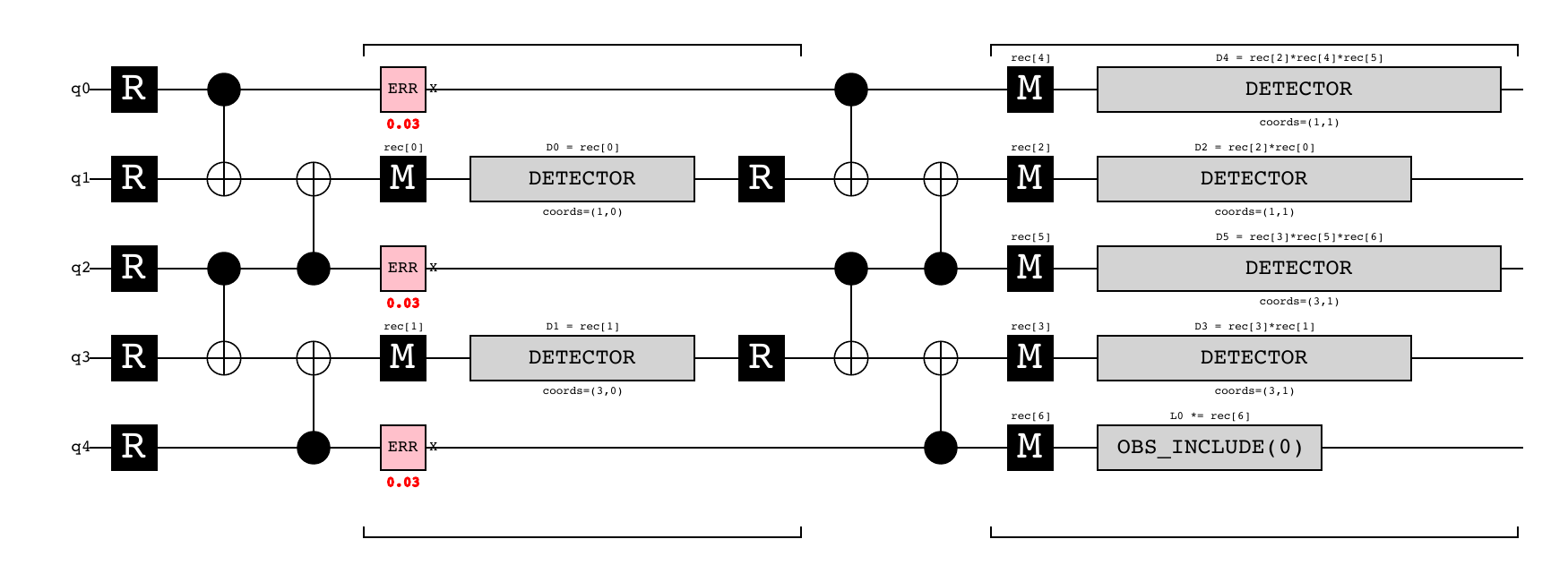}} \\
    \subfloat[\label{fig: stim_circuit_rc_pec}]{\includegraphics[width=0.5\textwidth]{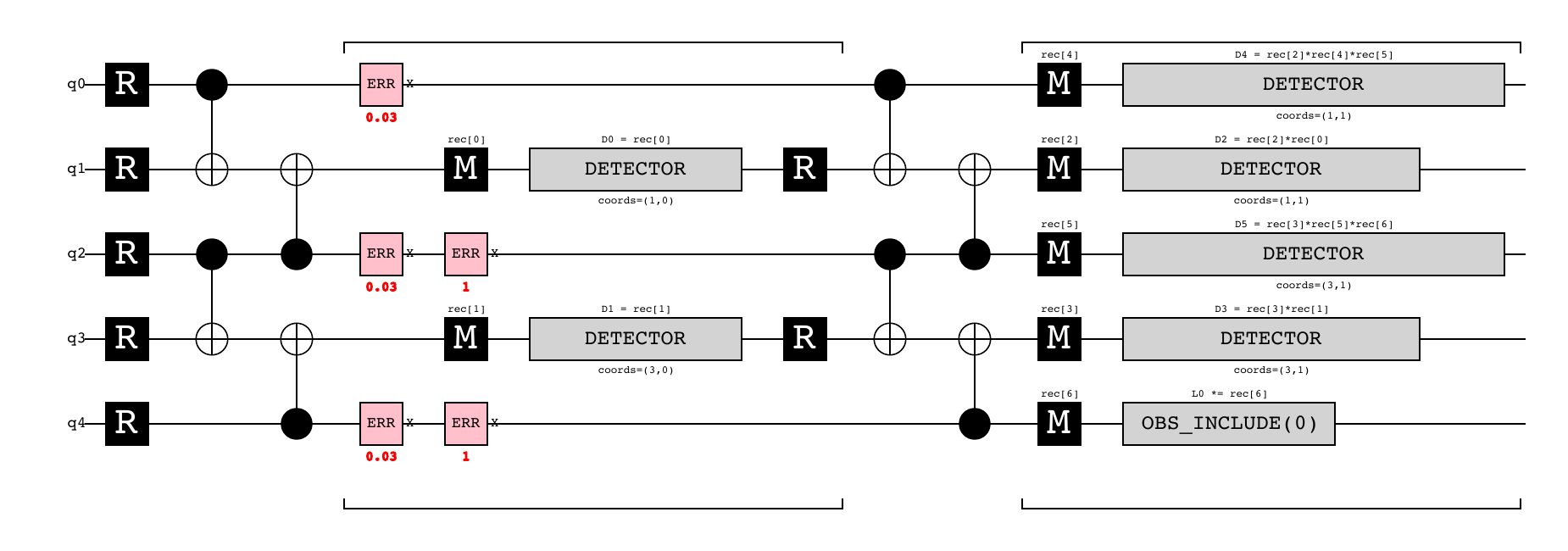}}

    \caption{Illustration of the distance-$3$ repetition code with bit-flip errors ($p=0.03$): (a) Unmitigated circuit (identity branch) (b) circuit with PEC (superbranch). The first round of stabiliser check here effectively projects the incoming state into the code space (when keeping track of the relevant correction). There are additional weight-$2$ bit-flips sampled from the superbranch. ``R'' stands for initialisation, ``M'' stands for measurement. The red boxes stand for noise locations and additional gates inserted by PEC.}
\label{fig: stim_circuit_rc}
\end{figure}

Using Stim~\cite{gidney2021stim}, we simulate the memory experiment of logical $\ket{0_{L}}$ under one round of code-capacity errors, followed by stabiliser measurements and decoding. Our scheme naturally extends to any fixed number of rounds (see \cref{subsec: two_equiv_pictures_qem_qec}). In the code-capacity model, we have single-qubit Pauli noise between rounds and assume perfect state preparation and measurement. We study the repetition code with bit-flip noise, and the rotated surface code with depolarising noise. For the bit-flip noise, we apply $X$ on each data qubit with probability $p$; for the depolarising noise, we have $X$, $Y$ or $Z$ errors occurring to each qubit with probability $p/3$. We consider distances $d\in\{3,5,7,9\}$. For each configuration $(\text{code}, d, p,\text{noise model})$, we run both the unmitigated (without PEC) and PEC experiments, and record $0$ or $1$ per shot, representing the absence or presence of logical error after the QEC recovery operation. We used a standard minimum weight perfect matching (MWPM) decoder, implemented in Pymatching~\cite{Higgott2025sparseblossom}. Exact circuit templates for the repetition code under bit-flip noise are given in \cref{fig: stim_circuit_rc} (see \cref{fig: stim_circuit_sc} for the surface code). Besides numerics, we have also written down the exact analytic expression of the logical error rate for the repetition code and semi-analytic expression of the logical error rate for the surface code in \cref{appendix: sec: branch_cond_rep_code_surface_code}. 

As discussed in \cref{subsec: inversion_channel}, the inverse channel we use targets all weight-$\omega=\lceil d/2\rceil$ errors. To improve the simulation efficiency, we stratify draws within each branch (see \cref{appendix: sec: strat_sampling})~\cite{chen2025fasterprobabilisticerrorcancellation, Dai2025StratifiedQPD}, which preserves the bias and results in tighter confidence intervals than naive Monte Carlo sampling.

\begin{figure*}
	\centering
	\subfloat[]{
    \includegraphics[width=0.495\linewidth]{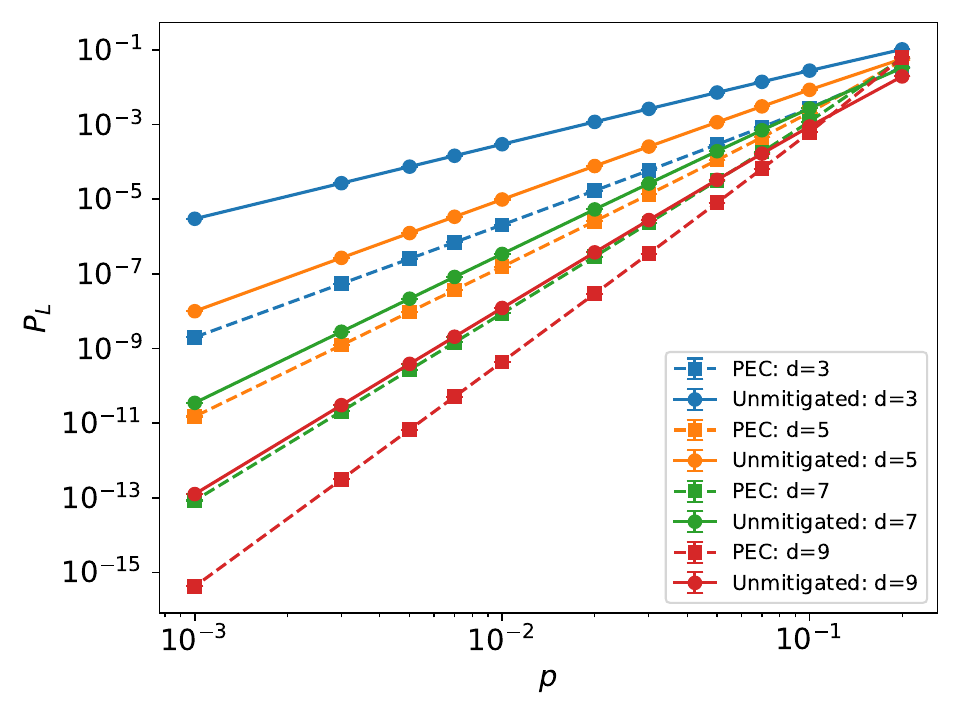}
    \label{fig:P_L_rc_x}
    }
    \subfloat[]{
   \includegraphics[width=0.495\linewidth]{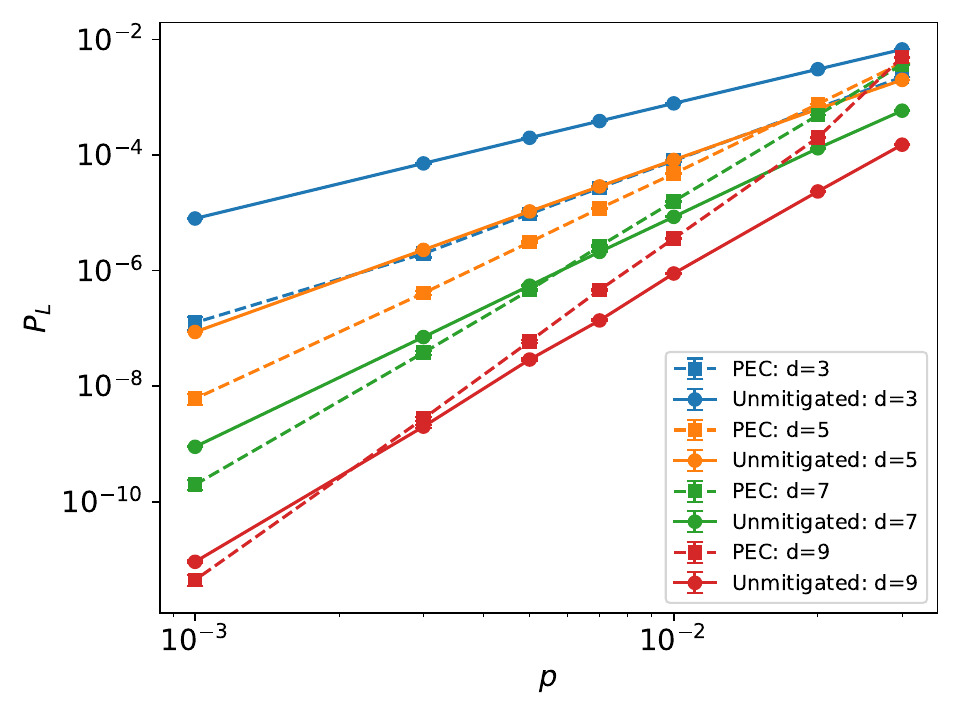}
   \label{fig:P_L_sc_depol} 
      }
    \caption{ Logical error rates with and without PEC against physical error rates for (a) repetition code under bit-flip noise (b) rotated surface code under depolarising noise. Note that we are actually plotting the absolute value of the logical error rate for the PEC curves here since the actual value is negative as described in \cref{sec:negative_log_err}.}
    \label{fig:P_L_plots}
\end{figure*}

For bit-flip repetition codes in bit-flip errors, the log-log plot of the logical error rates with PEC ($P_L^{\mathrm{PEC}}$) and without PEC ($P_L^{(0)}$) against the physical error rate ($p$) is shown in \cref{fig:P_L_rc_x}. For $d\in\{3,5,7,9\}$ the slopes for the unmitigated curves are $\{1.98,2.95,3.92,4.89\}$ while the PEC slopes are $\{3.11,4.11,5.12,6.13\}$, closely matching the predicted dominating order of error probability $\order{p^{\lceil d/2\rceil}} \xrightarrow{PEC} \order{p^{\lceil d/2\rceil+1}}$. The deviation of the slopes from exact integer values is likely due to the precise set-up of the decoder for borderline cases and contributions from higher-order error terms. Going beyond the slope of the curve and looking at the exact logical error rates achieved, we see that $P_L^{\mathrm{PEC}}(p;d) \lesssim P_L^{(0)}(p;d+2)$. Hence, applying PEC at the physical level of a repetition code, we can achieve the same error scaling as a larger repetition code with distance increased by 2, and a lower absolute logical error rates compared to the larger code. This improvement in the logical error rate is more pronounced at smaller code distances. 

The corresponding results for rotated surface codes in depolarising errors are shown in \cref{fig:P_L_sc_depol}. The fitted slopes are $\{1.98,2.96,3.93,4.90\}$ (unmitigated) versus $\{2.91,3.94,4.94,6.04\}$ (PEC) for $d=\{3,5,7,9\}$. Again, we see that the effective order of the uncorrectable errors is increased $\order{p^{\lceil d/2\rceil}} \xrightarrow{PEC} \order{p^{\lceil d/2\rceil+1}}$. However, even though PEC can achieve a logical error scaling that is the same as increasing the code distance by $2$, the exact logical error rate achieved follows $P_L^{\mathrm{PEC}}(p;d) \gtrsim P_L^{(0)}(p;d+2)$ in our simulation, which is the opposite of what happen for repetition code. 

There are two possible reasons for the difference between the logical error rate of the PEC case and the unmitigated code with a larger distance: $P_L^{\mathrm{PEC}}(p;d) $ vs $ P_L^{(0)}(p;d+2)$. First of all, the removal of all weight-$\lceil d/2\rceil$ errors using PEC will inevitably remove many error configurations that could be corrected through QEC. In other words, in addition to removing logical errors, PEC removes some error-free components. These error-free components from weight-$\lceil d/2\rceil$ errors, on the other hand, are perfectly preserved in distance-$(d+2)$ unmitigated codes. Thus, the distance-$d$ PEC code has a higher effective logical error rate due to the reduced error-free component compared to the distance-$(d+2)$ unmitigated code. Note that this will not affect the repetition code because it is non-degenerate and thus all weight-$\lceil d/2\rceil$ errors are damaging. This deviation should also be more pronounced at higher distances due to more ways to construct these error-free weight-$\lceil d/2\rceil$ errors, which is what we see in \cref{fig:P_L_sc_depol}.
Another possible reason for the deviation between $P_L^{\mathrm{PEC}}(p;d)$ $P_L^{(0)}(p;d+2)$ is the different number of error configurations leading to logical errors for both setups due to the different code sizes and the action of the PEC inverse channel. 

As seen above, in both repetition codes and surface codes, the improvements brought by PEC decrease with increased distance. Moreover, the qubit saving achieved by increasing the effective distance by $2$ is also most prominent at small distances. In our example, at distance $3$, the PEC repetition code can achieve $\sim 5$ times smaller logical error rate than distance-$5$ unmitigated repetition code while using $40\%$ fewer qubits ($3$ vs $5$). The distance-$3$ PEC surface code can achieve almost the same logical error rate as distance-$5$ unmitigated surface code while using $64\%$ fewer qubits ($9$ vs $25$).

The decreasing improvement brought by PEC with increased distance also lead to smaller error threshold for the PEC case. The error threshold with PEC is roughly $p_{\mathrm{th}}^{\mathrm{PEC}} \approx 0.19$ and $0.03$ for repetition codes and surface codes, respectively. These values are lower than the unmitigated error threshold $p_{\mathrm{th}} \approx 0.32$ and $0.12$ (these are obtained from linear extrapolation with the small-distance curves in our plot, thus differ from the theoretical value of $0.5$ and $0.15$). Further discussions of the relative sizes of these thresholds can be found in \cref{sec:error_threshold}. 

In the PEC plots, the error-mitigated logical error rates are actually negative, corresponding to negative contributions from the signed branch dominating. The source of this negativity in the logical error rate is further explained in \cref{sec:negative_log_err}. In practice, negative values are harmless in expectation value estimation since they simply mean flipping the sign of the bias without increasing its magnitude. In the case of applying to sampling problems, it can even be used to our advantage to remove some shot noise~\cite{liu2025quantumerrormitigationsampling}.

As mentioned in \cref{subsec: inversion_channel}, we need to operate at a physical error rate $p < p_{\mathrm{pole}}$ as given by \cref{eqn: pole} in order for the inverse channel in PEC to work effectively. For repetition codes, we have $N = d$ and $\omega = (d+1)/2$ (considering only odd $d$), and using Stirling's approximation, we know that $p_{\mathrm{pole}} > 0.2$, thus we are operating at a physical error rate safely below $p_{\mathrm{pole}}$. For surface codes, we have $N = d^2$ instead and the closest pole is for $d = 9$ which gives $p_{\mathrm{pole}} \approx 0.03$. This will lead to a deviation from the straight line at the $p = 0.02$ data point because the higher-order errors are no longer negligible near $p_{\mathrm{pole}}$ as discussed in \cref{subsec: inversion_channel}. More detailed comparisons between theory and simulation curves are provided in \cref{appendix: sec: more_result_plots}.

\section{Conclusions}
\label{sec: conclusions}

We have established a rigorous and general framework demonstrating that any linear QEM techniques applied to physical qubits are naturally compatible with quantum error correction (QEC). By exploiting the linearity of QEC, we proved that no modification to the decoder is required; the QEC process remains agnostic to the specific QEM technique applied at the physical layer. In many cases, the error-mitigated physical layer simply presents a cleaner effective channel to the QEC code.

Focusing on probabilistic error cancellation (PEC), we analytically proved and numerically verified that mitigating physical errors eliminates the leading-order $\order{p^{\lceil d/2\rceil}}$ logical error term. This effectively suppresses the logical error rate to $\order{p^{\lceil d/2\rceil + 1}}$, increasing the effective code distance by $2$. While this steepening of the error curve results in a lower error threshold compared to unmitigated QEC, it offers substantial performance gains in the low-error regime typical of early fault-tolerant devices.

The practical improvements achieved by this hybrid approach is most significant at smaller code distances, which is usually when the qubit overhead is the most critical constraint. Our simulations reveal that a distance-$3$ repetition code with PEC achieves a logical error rate $\sim 5$ times lower than a distance-5 unmitigated code while using $40\%$ fewer qubits. Similarly, a distance-3 surface code with PEC rivals the performance of a distance-5 unmitigated surface code with a $64\%$ reduction in qubit count. These results suggest that physical-level QEM is a powerful strategy for maximizing the computational power of near-term hardware with limited qubit budgets.

Looking forward, several avenues warrant further investigation. First, while our theoretical framework naturally extends to the circuit-level noise model, it will be interesting to test this numerically to further verify the robustness of the framework. Additionally, since our framework supports any linear QEM technique, exploring exact implementation with symmetry verification and virtual purification can also be interesting. Finally, while we demonstrated that no modification to the QEC process is required, it does not rule out the possibility of further adapting the QEC process (e.g. the decoder) to improve the logical performance. 

\section*{Acknowledgements}
The authors would like to acknowledge the use of the University of Oxford Advanced Research Computing (ARC) facility\cite{oxford_arc} in carrying out this work and specifically the facilities made available from the EPSRC QCS Hub grant (agreement No. EP/T001062/1). ZC is supported by the EPSRC quantum technologies career acceleration fellowship (UKRI1226) and EPSRC projects Robust and Reliable Quantum Computing (RoaRQ, EP/W032635/1).

\appendix
\section{Pole of inverse channel}\label{sec:pole}
From \cref{eqn: pole}, we know that the physical error rate that will lead to blow up of variance of the inverse channel is
\begin{align*}
    p_{\mathrm{pole}}=\frac{1}{\,1+\binom{N}{\omega}^{1/\omega}\,}
\end{align*}
Note that, since $\omega = \lceil d/2 \rceil$ and $N$ always increase with increased $d$ (and $N \geq d$), we know that $p_{\mathrm{pole}}$ will always decrease with increased $d$.

\subsection{Repetition Code}
For repetition code, we have $N = d$ and $\omega = (d+1)/2$ (considering only odd $d$). Taking the limit of large $d$ and using Stirling's approximation we have $p_{\mathrm{pole}}$ approaching $0.2$. Hence, we have $p_{\mathrm{pole}} > 0.2$ for all distances, which is far away from our simulated error rate of $< 0.1$. 

\subsection{Rotated Surface Code}
For rotated surface code, we have $N = d^2$ and $\omega = (d+1)/2$ (considering only odd $d$). Taking the limit of large $d$ and using Stirling's approximation, we have $p_{\mathrm{pole}} \approx (2ed)^{-1}$.

\subsection{Surface Code in Circuit Level Noise}
For rotated surface code in circuit level noise, $N$ will now be the number of error locations in the decoding graph, which scales as $N = d^3$. We again have $\omega = (d+1)/2$ (considering only odd $d$). Taking the limit of large $d$ and using Stirling's approximation we have $p_{\mathrm{pole}} \approx (2ed^2)^{-1}$

\section{PEC Error Threshold}\label{sec:error_threshold}
Assuming the case in which all of the PEC curves of different $d$ are straight lines and crossing over at the same threshold point $(p_{\mathrm{th}}^{\mathrm{PEC}}, P_{\mathrm{th}}^{\mathrm{PEC}})$, this means that they follows the following equation:
\begin{align}\label{eqn:log_err_pec}
    P_L^{\mathrm{PEC}}(p;d) = P_{\mathrm{th}}^{\mathrm{PEC}} \left(p/p_{\mathrm{th}}^{\mathrm{PEC}}\right)^{\lceil d/2\rceil+1} = D_{\lceil d/2\rceil+1}^{\mathrm{PEC}} p^{\lceil d/2\rceil+1}
\end{align}
where $D_{\lceil d/2\rceil+1}^{\mathrm{PEC}} = P_{\mathrm{th}}^{\mathrm{PEC}}{p_{\mathrm{th}}^{\mathrm{PEC}}}^{-\lceil d/2\rceil-1}$ is the number of error configurations with probability $\order{p^{-\lceil d/2\rceil-1}}$ that cause logical errors after applying PEC. 

Assuming the curves without PEC also follows a similar equation with the threshold value being $(p_{\mathrm{th}}, P_{\mathrm{th}})$:
\begin{align}\label{eqn:log_err_umiti}
    P_L^{(0)}(p;d) = P_{\mathrm{th}} \left(p/p_{\mathrm{th}}\right)^{\lceil d/2\rceil} = D_{\lceil d/2\rceil} p^{\lceil d/2\rceil}
\end{align}
where $D_{\lceil d/2\rceil} = P_{\mathrm{th}}p_{\mathrm{th}}^{-\lceil d/2\rceil}$ is the number of error configurations with probability $\order{p^{-\lceil d/2\rceil}}$ that cause logical errors. 

By dividing the expression of $D_{\lceil d/2\rceil+1}^{\mathrm{PEC}}$ with the expression of $D_{\lceil d/2\rceil}$, we have:
\begin{align*}
    p_{\mathrm{th}} 
    &= p_{\mathrm{th}}^{*}\left(\frac{p_{\mathrm{th}}}{p_{\mathrm{th}}^{\mathrm{PEC}}}\right)^{\lceil d/2\rceil + 1}\\
     p_{\mathrm{th}}^{\mathrm{PEC}} &= p_{\mathrm{th}}^{*}\left(\frac{p_{\mathrm{th}}}{p_{\mathrm{th}}^{\mathrm{PEC}}}\right)^{\lceil d/2\rceil}
\end{align*}
with 
\begin{align*}
    p_{\mathrm{th}}^{*} =  \frac{ D_{\lceil d/2\rceil}} {D_{\lceil d/2\rceil+1}^{\mathrm{PEC}}}\frac{P_{\mathrm{th}}^{\mathrm{PEC}}}{P_{\mathrm{th}} }.
\end{align*}
Hence, when $p_{\mathrm{th}} > p_{\mathrm{th}}^{\mathrm{PEC}}$, we have $p_{\mathrm{th}} > p_{\mathrm{th}}^{\mathrm{PEC}} > p_{\mathrm{th}}^*$. Similarly, when $p_{\mathrm{th}} < p_{\mathrm{th}}^{\mathrm{PEC}}$, we have $p_{\mathrm{th}} < p_{\mathrm{th}}^{\mathrm{PEC}} < p_{\mathrm{th}}^*$. 

Therefore, $p_{\mathrm{th}}^*$ is the critical value to compare against the threshold value of the unmitigated error threshold. When $p_{\mathrm{th}} > p_{\mathrm{th}}^*$, we have $p_{\mathrm{th}} > p_{\mathrm{th}}^{\mathrm{PEC}}$ and when $p_{\mathrm{th}} < p_{\mathrm{th}}^*$, we have $p_{\mathrm{th}} < p_{\mathrm{th}}^{\mathrm{PEC}}$. 

\section{Negative Logical Error Rate}\label{sec:negative_log_err}
Following arguments in \cref{sec: memory}, for the $\order{p^{\omega+1}}$ term in $\mathcal{F} \circ \mathcal{E}$, we have:
\begin{align}\label{eqn:wplus1_err}
    P_{\omega+1} \sum_{|K|=\omega+1} \mathcal{G}_{K} - \left(P_{1} \sum_{|K|=1} \mathcal{G}_{K}\right) \left(P_{\omega} \sum_{|K|=\omega} \mathcal{G}_{K}\right)
\end{align}
Assuming all error events containing above will lead to logical errors, the resultant logical error rate for the $\order{p^{\omega+1}}$ term will be
\begin{align}\label{eqn:wplus1_err_1}
    P_{\omega+1} \binom{N}{\omega+1} - P_1P_{\omega} N\binom{N}{\omega}
\end{align}
The ratio between the two term is:
\begin{align}
    \frac{P_{\omega+1} \binom{N}{\omega+1}}{P_1P_{\omega} N\binom{N}{\omega}} = \frac{N-\omega}{N (\omega + 1)(1-p)^N} = \frac{(1-\omega/N)/(\omega + 1)}{(1-p)^N}
\end{align}
Hence, \cref{eqn:wplus1_err_1} will be negative
\begin{align}
    (1-p)^N > \frac{1-\omega/N}{\omega + 1}
\end{align}
This is the case for our simulation where $Np \ll 1$. However, in practice for larger systems, the resultant logical error rate is more likely to be positive. 

\section{Application of the PEC Framework to Bit-Flip and Depolarising Noise} \label{appendix: sec: derivation_err_mit_channel}

In \cref{sec: memory}, we developed a general framework of PEC to memory experiment under code-capacity model with a general error channel $\mathcal{G}$.

In this appendix, we present a specific example of applying this general framework to the memory experiment under bit-flip noise: We explicitly derive the closed-form expressions of (a) the effective error channel $\mathcal{F} \circ \mathcal{E}$ (\cref{eqn: effective_err_channel_memory}) and (b) the error-mitigated logical error rate, $P^{\mathrm{PEC}}_{L}$. Furthermore, we show the suppression of $\order{p^{\omega}}$ terms in the error-mitigated logical error rate under bit-flip noise model. In addition, we comment that the same argument trivially applies to the depolarising noise.

\subsection{Effective Error Channel under the Bit-Flip noise}

Consider $N$ data qubits of a QEC code of distance $d$ in a memory experiment. In the code-capacity model, we assume perfect measurement without any error, and the data qubits are subject to bit-flip errors with the probability, $p$, in between the stabiliser checks. 

The error channel in \cref{eqn: pauli_error_channel_weight_w} can be written as follows:

\begin{equation} \label{eqn: bit_flip_channel_weight_w}
    \mathcal{E} = \Bigr[\bigl (1-p \bigr )\mathcal{I} + p\mathcal{X}\Bigl]^{\otimes N} =  \sum^{N}_{k=0}P_{k} \sum_{|K|=k} \mathcal{X}_{K},
\end{equation}where $\mathcal{X}$ is the superoperator of Pauli $X$, i.e. $\mathcal{X}(\cdot) = X(\cdot)X$, $K\subseteq [N]$ is a support set denoting the locations of bit-flips, and $P_{k}$ is the probability having a weight-$k$ error, i.e. $P_{k} \equiv p^{k}(1-p)^{N-k}$ with $k=|K|$ (see \cref{eqn: weight_k_error_prob}).

For the bit-flip noise channel, the approximate inverse channel in \cref{eqn: approx_inv_channel} becomes

\begin{equation} \label{eqn: approx_inv_channel_bit_flip}
    \mathcal{F} = \frac{1}{A}\left(P_{0}\mathcal{I} - P_{\omega}\sum_{|S| = \omega}\mathcal{X}_{S}\right),
\end{equation}where the sum runs over all supports of weight-$\omega$ errors, i.e. $|S| = \omega$, $P_{0} \equiv (1-p)^{N}$, $P_{\omega} \equiv p^{\omega}(1-p)^{N-\omega}$, and $A \equiv P_{0} - \Binom{N}{\omega}P_{\omega}$, i.e. the number of all weight-$\omega$ errors.

Composing the bit-flip channel and the approximate inverse channel, results in the following effective error channel:

\begin{equation}
\label{eqn: composite_channel_interm}
\begin{aligned}
\mathcal{F}\circ\mathcal{E}
  &= \frac{1}{A}\sum^{N}_{k=0} P_{k} \sum_{|K|=k}\Big[
      P_{0}\mathcal{X}_{K} - P_{\omega} \sum_{|S|=\omega}\,\mathcal{X}_{S}\mathcal{X}_{K}
    \Big].
\end{aligned}
\end{equation}Since $\mathcal{X}_{S}\mathcal{X}_{K} = \mathcal{X}_{S \oplus K}$, one may reduce the second term in the bracket as a sum over $V=S \oplus K$. For a particular output support $V$ with $|V|=v$ and a fixed support $S$ with $|S|=\omega$, let $t=|S \cap V|$. Then, $K=S \oplus V$ satisfies $|K| =\omega + v - 2t$. For a fixed $V$, the number of support sets $S$ of weight-$\omega$ that meet $V$ in exactly $t$ positions is given by:

\begin{equation}
    \left \vert \, \{S: |S| = \omega, \: |S \cap V| = t\}\, \right \vert = \Binom{v}{t}\Binom{N-v}{\omega-t}.
\end{equation}Each such $S$ contributes a term with probability $p^{|K|} (1-p)^{N-|K|} = p^{\omega + v - 2t}(1-p)^{N-(\omega+v-2t)}$. Thus, the exact expression of the resultant channel after PEC is
\begin{align}
\mathcal{F} \circ \mathcal{E}
  &= \sum_{k=0}^{N} c_k(p)\sum_{|K|=k}\mathcal{X}_K, 
\label{eqn: composite_quantum_channel}
\end{align}with the coefficients for the given weight, $|V|=v$:

\begin{equation}\label{eqn: cvp}
\begin{split}
&c_k(p)   \\
&=\frac{1}{A}\bigg[ P_0 P_k  - P_\omega \sum_{t=\max(0,\omega+k-N)}^{\min(\omega,k)}
     \binom{k}{t}\binom{N-k}{\omega-t} P_{\omega+k-2t}\bigg].
\end{split}
\end{equation}

\subsection{Logical Error Rate under Bit-flip Noise} \label{sec:bit_flip_log_error_rate}

Note that $c_\omega(p) \neq 0$, i.e. not all weight-$\omega$ errors are removed. However, their probabilities are suppressed to $\order{p^{\omega+1}}$.

Since the code is of distance $d$, looking at \cref{eqn:pec_qem_qec_2}, we have $\pbra{1_L}\mathcal{R}\mathcal{X}_K\pket{0_L} = 0$ for all $\mathcal{X}_K$ with weight smaller than $\omega = \lceil d/2\rceil$, i.e. all of such errors can be perfectly corrected. For errors with a given weight $k\geq \omega$, we will use $D_{k}$ to denote the number of weight-$k$ error patterns on $N$ data qubits that lead to an uncorrectable logical error, i.e. $\pbra{1_L}\mathcal{R}\mathcal{X}_K\pket{0_L} = 1$ for these errors. Note that $D_k \leq \binom{N}{k}$.

In this way, using  \cref{eqn: composite_quantum_channel}, we then have:
\begin{equation}
P^{\mathrm{PEC}}_L(p;d)=\sum_{k\geq \omega}^N c_k(p)\,D_{k},
\label{eqn: log_err_rate_comp_channel}
\end{equation}
which is $\sim \order{p^{\omega+1}}$.
All dependence on the code geometry and decoder resides in the integers $D_{k}$, whose evaluation and use in theoretical predictions are detailed in \cref{appendix: sec: branch_cond_rep_code_surface_code}. The effect of PEC enters solely through the coefficients $c_k(p)$.

\subsection{Cancellation of $p^{\omega}$ terms in the Error-mitigated Logical Error Rate}\label{appendix: subsec: p_cancellation_branch_conditioned}

We now prove the cancellation of $\order{p^{\omega}}$ terms in the error-mitigated logical error rate, i.e. $P^{\mathrm{PEC}}_{L}(p;d) = \order{p^{\omega+1}}$. Here, we adapt the branch-conditioned picture introduced in \cref{subsec: log_err_rate} because it has a clearer connection to the numerical simulations present in this paper (see in \cref{appendix: sec: branch_cond_rep_code_surface_code}); the logical error rate in \cref{eqn: log_err_rate_branch_cond_no_gain} is equivalent to the operational definition of the Monte Carlo estimator of PEC used in practice (see \cref{appendix: sec: operational_definition}).

The logical error rate of the identity component is given by:

\begin{align} 
    P^{(0)}_{L}(p) &= \sum_{k\ge \omega} D_k\,p^k(1-p)^{N-k} \label{eqn:fail_prob_identity_branch_cond} \\
    &= D_{\omega}p^{\omega}(1-p)^{N-\omega} + D_{\omega+1}p^{\omega +1} (1-p)^{N-(\omega+1)} + \order{p^{\omega+2}}, \nonumber
\end{align}where $D_{k}$ counts the number of the weight-$k$ error patterns that lead to a logical error.

The logical error rate of the superbranch components is given by:

\begin{equation} \label{eqn: fail_prob_super_branch_branch_cond}
\begin{aligned}
    P^{(\omega)}_{L}(p) &= f_{0}(1-p)^{N} + \underbrace{\left((1-f_0)\bar{\kappa} - f_{0}\bar{\mu}\right)}_{s_{1}}p(1-p)^{N-1} + \order{p^{2}}, \\
    &= f_{0}(1-p)^{N} +s_{1}p(1-p)^{N-1} + \order{p^{2}} \\
\end{aligned}
\end{equation}where $f_0$ is the probability that the weight-$\omega$ errors introduced by the PEC will lead to a logical error, i.e. $f_{0} = D_{\omega}/\Binom{N}{\omega}$ for bit-flip noise, $\bar{\kappa}$ is the average number of single-qubit flips that complete a non-failing error patterns into a failing pattern (averaged over $\Binom{N}{\omega}-D_{\omega}$ non-failing pairs), $\bar{\mu}$ is the average number of single-qubit flips that neutralize a failing pair (averaged over the $D_{\omega}$ failing pairs). We defined $s_{1} \equiv (1-f_{0})\bar{\kappa} - f_{0}\bar{\mu}$ to capture the coefficient of order-$p$ in $P^{(\omega)}_{L}$. Note that $f_{0}$ depends on the specifics of the quantum error correcting code and the approximate inverse channel used.

Recall the logical error rate in \cref{eqn: log_err_rate_branch_cond_no_gain}. Considering its numerator, the identity component at order $p^{\omega}$ comes from weight-$\omega$ physical errors:
\[[p^{\omega}](P_{0} P^{(0)}_{L})=P_{0}\,D_{\omega}\,p^{\omega}(1-p)^{N-\omega}=P_{0}P_{\omega} D_{\omega}.\]
In the superbranch component, the $P_{\omega}$ already contributes $p^{\omega}$; an order-$p^{\omega}$ term occurs only for the lowest order term, i.e. $f_{0}(1-p)^{N}$, where the inserted PEC gates $S$ is a failing weight-$\omega$ error. So,
\[[p^{\omega}]\big(-\binom{N}{\omega}P_{\omega} P_S\big)=-P_{0}P_{\omega} \binom{N}{\omega}f_{0} = -P_{0}P_{\omega}D_{\omega}.\] Thus, the sum of the two terms is cancelled, i.e.

\[[p^{\omega}](P^{\mathrm{PEC}}_{L} (p;d))=0.\]

In \cref{appendix: sec: branch_cond_rep_code_surface_code}, we apply the theory developed in this section to two specific QEC codes: repetition code and rotated surface code.

\subsection{Comments on Depolarising Noise}
\label{appendix: subsec: depolarising_noise}

We comment that the cancellation of $\order{p^{\omega}}$ terms in the error-mitigated logical error rate also holds for the depolarising noise with changes only to the approximate inverse channel to target all weight-$\omega$ errors including both bit-flips and phase-flips. 

For $N$ data qubits subject to depolarising noise, the error channel in \cref{eqn: pauli_error_channel_weight_w} can be written as follows:

\begin{equation} \label{eqn: depol_channel_full}
    \mathcal{E} = \Bigr[\bigl (1-p \bigr )\mathcal{I} + \frac{p}{3}\left(\mathcal{X} + \mathcal{Y} + \mathcal{Z}\right)\Bigl]^{\otimes N},
\end{equation}where the curly letters are superoperators of the corresponding Pauli operators, e.g. $\mathcal{X}(\cdot) =X(\cdot)X$. The weight-$\omega$ contribution of \cref{eqn: depol_channel_full} is

\begin{equation}
\mathcal{S}_\omega:=\frac{1}{C}\sum_{|S|=\omega}\;\sum_{\mathbf{a}\in\{X,Y,Z\}^S}\mathcal{P}_{S,\mathbf{a}},
\end{equation}where $S \subset [N]$ is a support set of weight-$\omega$ errors ($|S|=\omega$), denoting the error locations; $\mathcal{P}_{S,\mathbf{a}}$ defines the Pauli channel, i.e.
$\mathcal{P}_{S,\mathbf{a}}(\rho)=P_{S,\mathbf{a}}\rho P_{S,\mathbf{a}}$ where $P_{S,\mathbf{a}}$ applies letter $a_j$ to site $j\in S$ and $I$ elsewhere; $C$ is the number of distinct weight-$\omega$ Pauli strings, $C :=\binom{N}{\omega}\,3^\omega$. 

While the total number of weight-$\omega$ errors increased by a factor of $3^{\omega}$, the probability of having a weight-$\omega$ error decreased by the same factor, i.e. $\bigl(\frac{p}{3}\bigr)^{\omega}(1-p)^{N-\omega}$. Thus, the approximate inverse channel, targeting all weight-$\omega$ errors, becomes
\begin{align}
\mathcal{F}
&=\frac{1}{A}\Bigl({P_{0}\mathcal{I}-\binom{N}{\omega}3^{\omega} \cdot\Bigl(\frac{p}{3}\Bigr)^{\omega}(1-p)^{N-\omega} \mathcal{S}_\omega}\Bigr) \\
&= \frac{1}{A} \Bigl(P_{0}\mathcal{I} - \Binom{N}{\omega}P_{\omega}S_{\omega}\Bigr),
\label{eqn: approx_inv_channel_depol}
\end{align}where $P_{0} = (1-p)^{N}$, $P_{\omega} = p^{\omega}(1-p)^{N-\omega}$; $\mathcal{I}$ and $\mathcal{S_{\omega}}$ lead to identity and superbranch components in logical error rates respectively, resulting in the same logical error rate in \cref{eqn: log_err_rate_branch_cond_no_gain}. Following the arguments in \cref{appendix: subsec: p_cancellation_branch_conditioned}, we can see the $\order{p^{\omega}}$ term is cancelled. Thus, the cancellation of weight-
$\omega$ contributions persists exactly under depolarising noise, demonstrating that our scheme only depends on the linearity of QEC and PEC.

\begin{figure}[htbp]
    \centering
    \subfloat[\label{fig: stim_circuit_sc_vanilla}]{\includegraphics[width=0.5\textwidth]{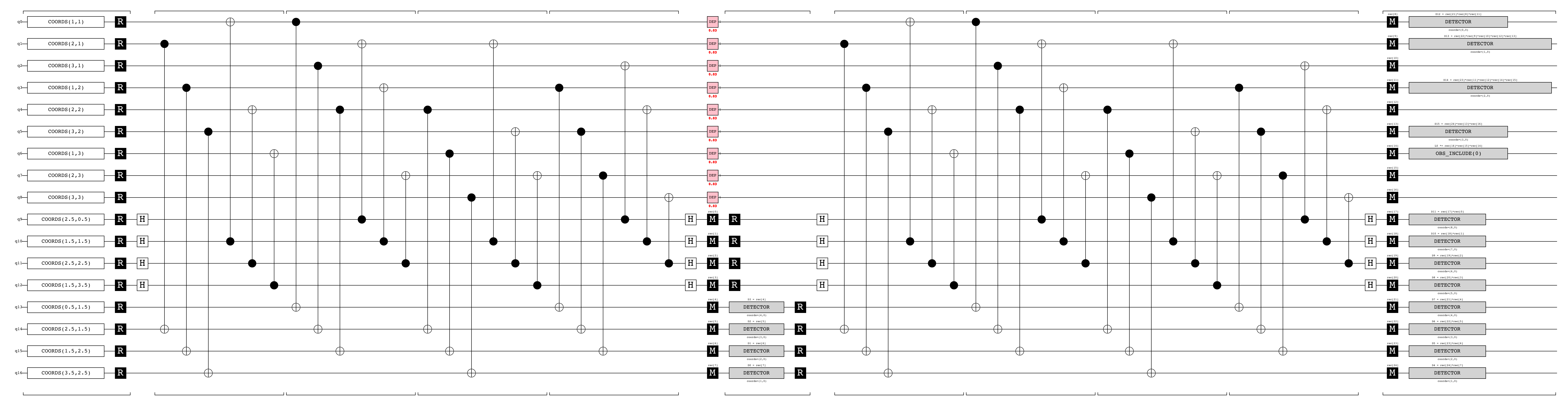}} \\
    \subfloat[\label{fig: stim_circuit_sc_pec}]{\includegraphics[width=0.5\textwidth]{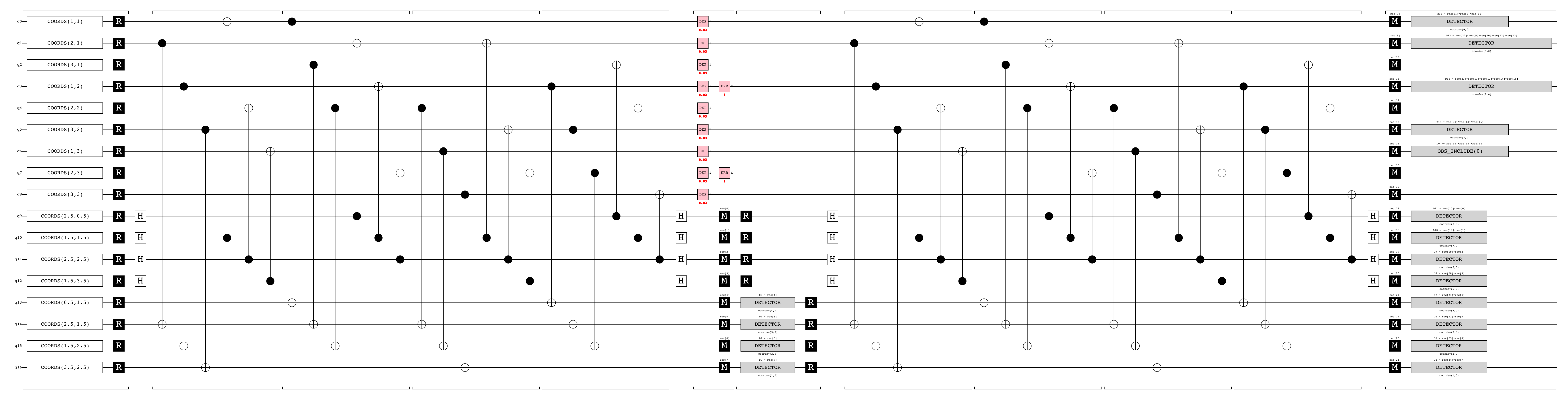}}

    \caption{Illustration of the distance-$3$ rotated surface code with depolarising errors ($p=0.03$): (a) Unmitigated circuit (identity branch) (b) circuit with PEC (superbranch). There are additional weight-$2$ errors sampled from the superbranch.}
\label{fig: stim_circuit_sc}
\end{figure}

\section{Operational Definition of PEC Logical Error Rate}
\label{appendix: sec: operational_definition}

This appendix derives the Monte-Carlo estimator of the error-mitigated logical error rate and shows that it is equivalent to the error-mitigated logical error rate in \cref{eqn: log_err_rate_branch_cond_no_gain} under linearity of QEC. Furthermore, we derive the sampling overhead introduced by the PEC, which will be further reduced by stratified sampling introduced in \cref{appendix: sec: strat_sampling}.

In practice, PEC is executed by the Monte Carlo sampling: In each shot, one runs a noisy circuit with the error channel $\mathcal{E}$ mitigated by the channel $\mathcal{B}_{b}$ drawn from the quasi-probability decomposition of the approximate inverse channel. Then, the QEC recovery channel applies the correction based on the syndrome of each shot. The final measurement is performed to determine the presence or absence of the logical error. The probability of sampling a branch index $b \in \{1,..., B\}$ is given by\cite{Cai_2023}:

\begin{equation}
    \pi(i=b) = |\beta_{b}|/\norm{\beta}_{1},
\end{equation}where the denominator is the L1 norm of the quasi probability decomposition, i.e. $\norm{\beta}_{1} = \sum_{b} |\beta_{b}|$. With the approximate inverse channel given by \cref{eqn: approx_inv_channel}, PEC samples the identity branch ($b=0$) and superbranch with the following probabilities:

\begin{align}
    \pi(0) &= \frac{P_{0}}{A}\nonumber \\
    \pi(i\neq0) &= \Binom{N}{\omega}\frac{P_{\omega}}{A}, \nonumber
\end{align} as the L1 norm of the quasi probability decomposition is 

\begin{align}
    \norm{\beta}_{1} = \sum_{b} |\beta_{b}| = \frac{P_{0} + \Binom{N}{\omega}P_{\omega}}{|A|},
\end{align}where $A = P_{0}-\Binom{N}{\omega}P_{\omega}$. The sampling overhead due to the PEC scales as the L1 norm squared, i.e. $\sim ||\beta||^{2}_{1} = \Bigl(\frac{P_{0}+\Binom{N}{\omega}P_{\omega}}{P_{0}-\Binom{N}{\omega}P_{\omega}}\Bigr)^{2}$.

The error-mitigated logical error rate is operationally defined by the average of measurement outcomes, denoted by $f \in \{0,1\}$, weighted by the sign of the weight $\beta_b$ and the L1 norm of the quasi probability decomposition over many such circuit runs:

\begin{equation}
P^{\mathrm{PEC}}_{L}\equiv\mathbb{E}[sign(\beta_{b})||\beta||_{1}f].
\label{eqn: log_err_rate_operational}
\end{equation}

We further rearrange the error-mitigated logical error rate in \cref{eqn: log_err_rate_branch_cond_no_gain} to prove the equivalence:

\begin{align}
P^{\mathrm{PEC}}_{L}
&= \norm{\beta}_{1}\Big( \pi(i=0)\,P^{(0)}_{L} \;-\; \pi(S)\,P^{(\omega)}_{L}\Big) \nonumber \\
&= \norm{\beta}_{1}\Big( \pi(I)\mathbb{E}_{f \sim P^{(0)}_{L}}[f] \;-\; \pi(S)\,\mathbb{E}_{f \sim P^{(\omega)}_{L}}[f]\Big) \nonumber \\
&= \norm{\beta}_{1}\bigl(\pi(I)\mathbb{E}_{f \sim P^{(0)}_{L}}[sign(\beta_{0})f] \\
& \qquad + \pi(S)\,\mathbb{E}_{f \sim P^{(\omega)}_{L}}[sign(\beta_{i \neq 0})f]\bigr) \nonumber \\
&= \mathbb E[sign(\beta_{b})\norm{\beta}_{1}f].
\end{align}We retrieve the operational definition of the error-mitigated logical error rate. This establishes rigorously that the shot-by-shot PEC implementation used in simulations corresponds exactly to the analytic expression of the error-mitigated logical error rate in \cref{eqn: log_err_rate_branch_cond_no_gain}.

\section{Logical Error Rates of the Repetition Code and Rotated Surface Code}
\label{appendix: sec: branch_cond_rep_code_surface_code}

In this section, we present analytic formulas for the repetition code and semi-analytic methods for the surface code to evaluate the logical error rates of the identity component $P^{(0)}_{L}$ and the superbranch component $P^{(\omega)}_{L}$. By combining them using \cref{eqn: log_err_rate_branch_cond_no_gain}, we obtained the theoretical predictions of the error-mitigated logical error rates of the repetition code in \cref{fig:P_L_rc_x}, \cref{fig: rc_xerror}, and the rotated surface code in \cref{fig:P_L_plots}, \cref{fig: sc_xerror}, and \cref{fig: sc_depol1_error}.

\subsection{Repetition Code} \label{appendix: subsec: branch_cond_rep_code}

For the repetition code, we can analytically derive the exact logical error rate as in \cref{eqn: log_err_rate_branch_cond}. The logical error rate of the identity component, i.e. the unmitigated logical error rate, is given by the probability of having errors whose weights are more than or equal to  $\omega \equiv \lceil\frac{d}{2}\rceil$:

\begin{equation}
    \label{eqn: PI_rep_code_analytic}
    P^{(0)}_{L}(p) = \sum^{N}_{k=\omega} \Binom{N}{k} p^{k}(1-p)^{N-k}.
\end{equation}On the other hand, the logical error rate of the superbranch component should account for the gates of weight-$\omega$ inserted by PEC. Let $u$ be the number of errors overlapping with the inserted gates $u\leq\omega$ and let $r$ be the number of extra errors, $r\leq N-\omega$. Note that the probability of a bit-flip error for the overlapping errors is $1-p$ due to the inserted PEC gates. Then, the logical error rate of the superbranch component is given by

\begin{equation}
\begin{aligned}
P^{(\omega)}_L(p)
&= \sum_{u=0}^{\omega} \sum_{r=0}^{N-\omega}
    \underbrace{\binom{\omega}{u} (1-p)^\omega}_{\text{overlaps}} p^{\,\omega-u}  \\
    &\times \qquad \underbrace{\binom{N-\omega}{r} p^{\,r} (1-p)^{(N-\omega)-r}}_{\text{extras}} \mathbf{1}_{\{u + r \ge \omega\}}\\[6pt]
&= \sum_{u=0}^{\omega} 
    \binom{\omega}{u} (1-p)^u p^{\,\omega-u} \\
    &\times \sum_{r=\max(0,\,\omega - u)}^{N-\omega}
    \binom{N-\omega}{r} p^{\,r} (1-p)^{(N-\omega)-r},
\end{aligned}
\label{eq:PS_rep_code_analytic}
\end{equation}where the indicator condition, $\mathbf{1}_{\{u + r \ge \omega\}}$, means that the inner sum over $r$ only includes terms satisfying $\omega - u \leq r$. From \cref{eqn: log_err_rate_branch_cond}, we could obtain the error-mitigated logical error rate analytically.

\subsubsection{Example: d=3}

We present $d=3$ repetition code as an example. For the identity branch, the logical error occurs for all weight-$2$ and weight-$3$ errors, resulting in the unmitigated logical error rate as follows:

\begin{align}
    P^{(0)}_{L}(p) &= \Binom{3}{2}p^{2}(1-p) + p^{3} = 3p^{2} - 2p^{3}. \nonumber
\end{align}

On the other hand, the superbranch inserts a weight-$2$ error on top of the error from the bitflip channel. The logical error occurs when the composite error becomes weight-$2$ or weight-$3$. This leads to the logical error rate of the superbranch component as follows:

\begin{align}
    P^{(\omega)}_{L}(p) &= (1-p)^{3} + (1-p)^{2}p + 2(1-p)p^{2} \nonumber \\
    &= 1 - 2p + 3p^{2} -2p^{3} \nonumber
\end{align}Combining them in \cref{eqn: log_err_rate_branch_cond}, one can obtain the following exact expression for the logical error rate:

\begin{equation}
\begin{aligned}
    P^{\mathrm{PEC}}_{L}(p; d=3) &= 3p^{2} - 2p^{3} - \frac{3p^{2}}{1-2p-2p^{2}} \times (1-2p) \\
    &= -2p^{3} + \order{p^{4}}.
\end{aligned}
\end{equation}The $p^{2}$ term is cancelled as expected.

\subsection{Rotated Surface Code} \label{appendix: subsec: branch_cond_surface_code}

Unlike repetition code, it is non-trivial to evaluate the component-wise logical error rates fully analytically. This is due to the difficulty in evaluating $D_{k}$'s, i.e. the number of weight-$k$ error patterns that lead to a logical error, with pure combinatorics; the exact values of $D_{k}$ depends on the specifics of the chosen decoder. In this section, we explain our semi-analytic method to predict the error-mitigated logical error rates.

\begin{figure*}
	\centering
    \subfloat[\label{fig: PI_sc_x_error}]{\includegraphics[width=0.5\textwidth]{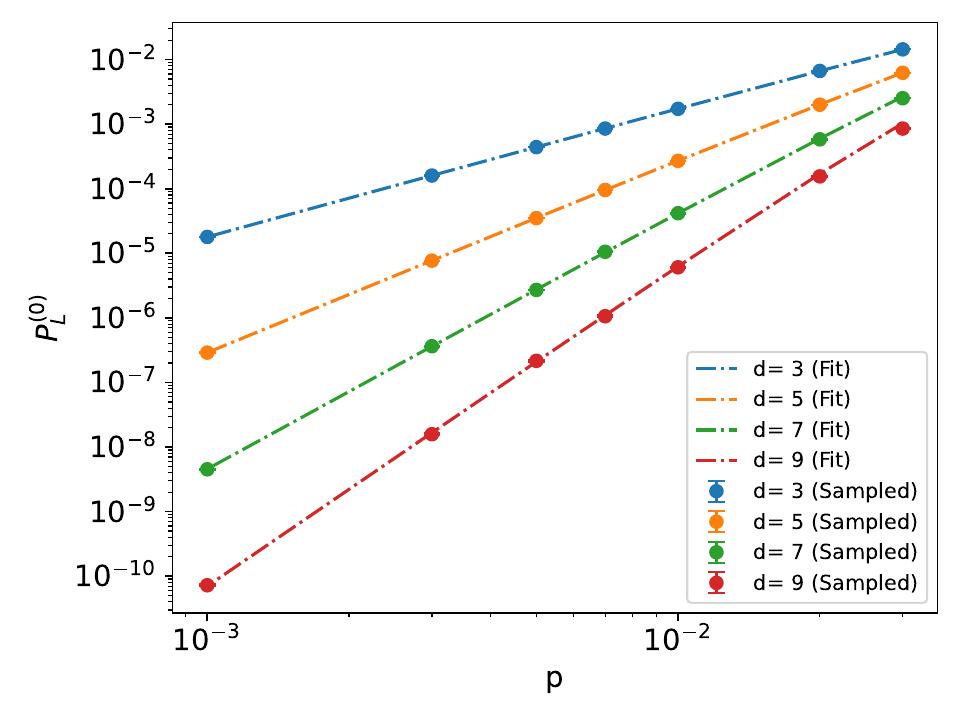}}
    ~
    \subfloat[\label{fig: PS_sc_x_error}]{\includegraphics[width=0.5\textwidth]{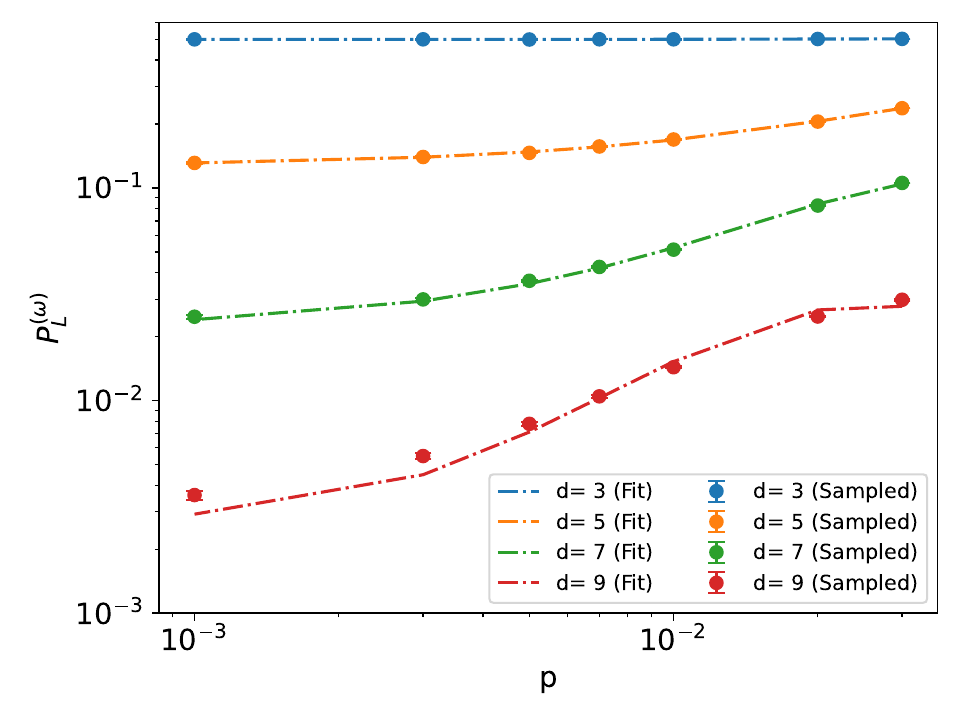}}
    \hfill
    
    \subfloat[\label{fig: PI_sc_depol1_error}]{\includegraphics[width=0.5\textwidth]{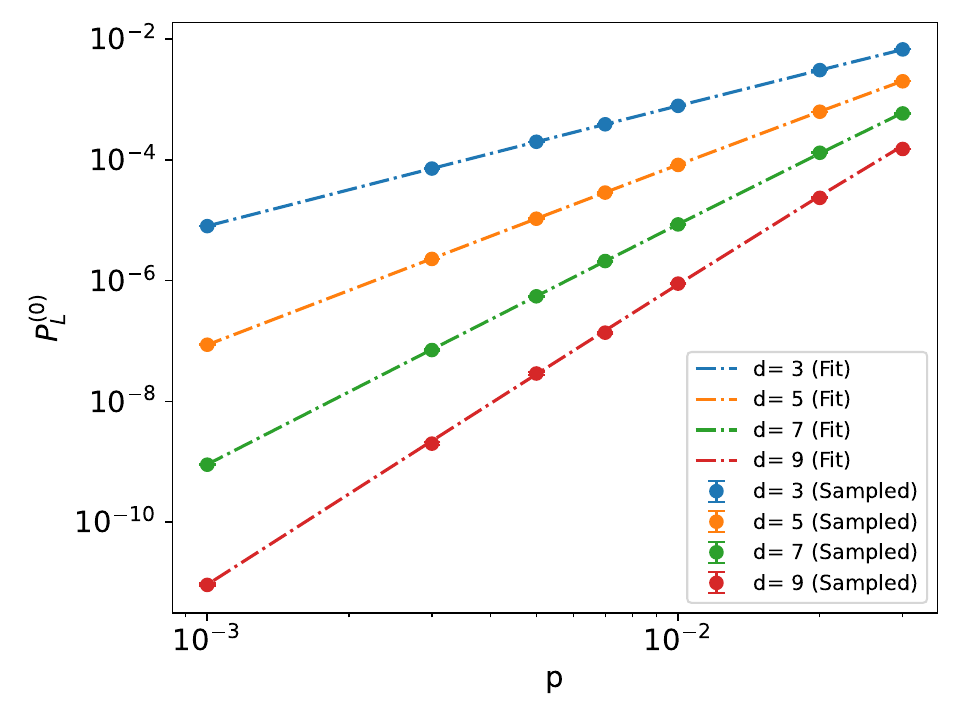}}
    ~
    \subfloat[\label{fig: PS_sc_depol1_error}]{\includegraphics[width=0.5\textwidth]{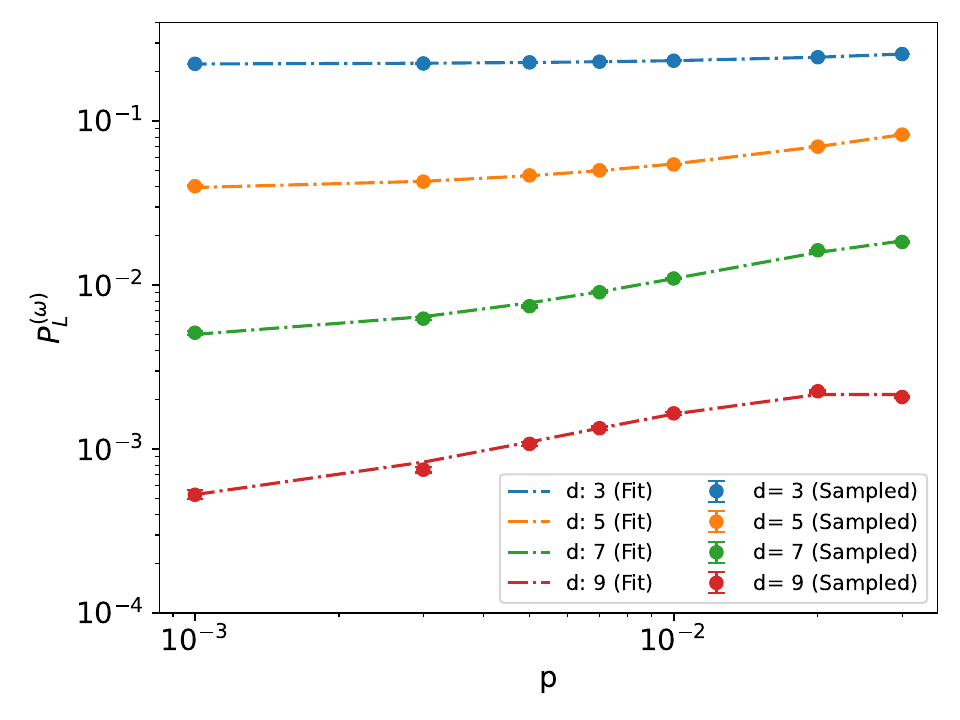}}
	
    \caption{The sampled (dots) and predicted/fitted (dotted lines) branch-conditioned logical error rates: (a,b) for bit-flip noise and (c,d) for the depolarising noise. The logical error rate of the identity branch, $P^{(0)}_{L}$, was given by its truncated series expression(\cref{eqn: PI_surface_branch_series}) with $D_{k}$'s directly obtained from the decoder. The logical error rate of the superbranch was estimated by fitting $s_{1}, \: s_{2}, \: s_{3}$ in its truncated series expression (\cref{eqn: PS_surface_branch_series}) on the sampled $P^{(\omega)}_{L}$.}
    \label{fig:PI_PS_sc}
\end{figure*}

Recall the unmitigated logical error rate in \cref{eqn:fail_prob_identity_branch_cond}. We truncate the series up to the third lowest order:

\begin{equation}
P^{(0)}_{L}(p)\approx\sum^{\omega+3}_{k=\omega} D_{\omega}\,p^k(1-p)^{N-k}.
\label{eqn: PI_surface_branch_series}
\end{equation}We determine $D_k$ by explicitly enumerating all weight-\(k\) errors and decode each pattern. For $d\in\{3,5,7\}$ we compute $D_\omega,\ldots,D_{\omega+3}$ exactly; for $d=9$ we compute \(D_\omega,D_{\omega+1}\) exactly and estimate \(D_{\omega+2}, D_{\omega+3}\) by Monte-Carlo with $10$M shots each. Note that truncating \cref{eqn:fail_prob_identity_branch_cond} at \(k=\omega+3\) faithfully reproduces the sampled identity branch logical error rates \(P^{(0)}_{L}\) across all distances (see \cref{fig: PI_sc_x_error} and \cref{fig: PI_sc_depol1_error}).

In the superbranch component, we truncate the superbranch logical error rate in \cref{eqn: fail_prob_super_branch_branch_cond} up to the third order:
\begin{equation}
\begin{aligned}
P^{(\omega)}_L(p)&\approx f_0(1-p)^N+s_1\,p(1-p)^{N-1}\\
 &+s_2\,p^2(1-p)^{N-2}+s_3\,p^3(1-p)^{N-3},
\end{aligned}
\label{eqn: PS_surface_branch_series}
\end{equation}
with $f_0=D_\omega/\binom{N}{\omega}$ fixed by the explicit enumeration of the weight-$\omega$ error patterns obtained previously. The coefficients $s_i$'s are obtained by least-squares fits directly to the sampled $P^{(\omega)}_L$. The fitted $s_{i}$'s are reported in \cref{tab:fs-bitflip-new} and \cref{tab:fs-depol-new}. Keeping terms up to $k=3$ suffices for the $p$-range explored (see \cref{fig: PS_sc_x_error} and \cref{fig: PS_sc_depol1_error}).

For depolarising noise, if $D_t^{(X)}$ counts the $X$-only failing errors of weight $t$, then the number of failing Pauli strings of total weight $k$ is
\begin{equation}
D_k^{(\mathrm{dep})}=\sum_{t=\omega}^{k} 2^{t}\,D_t^{(X)}\,\binom{N-t}{k-t},
\label{eqn: Aw_dep_mapping}
\end{equation}
since $Y$ acts like $X$ for the memory experiment of $\ket{0_{L}}$ and $Z$ acts like $I$. We plug $D_k^{(\mathrm{dep})}$ into \eqref{eqn: PS_surface_branch_series} and fit $P^{(\omega)}_L$ as in \eqref{eqn: PS_surface_branch_series}. 

Throughout, we used \cref{eqn: PI_surface_branch_series} and \cref{eqn: PS_surface_branch_series} to make the theoretical predictions of the error-mitigated error rates. These expressions enable direct evaluation of theoretical PEC logical-error curves, allowing quantitative comparison with the simulation data presented in \cref{fig: rc_xerror}, \cref{fig: sc_xerror}, and \cref{fig: sc_depol1_error}.

\begin{table}[h]
\centering
\caption{\textbf{Superbranch fit coefficients $(s_1,s_2,s_3)$ for the bit-flip channel} in \eqref{eqn: PS_surface_branch_series}. $f_0=D_\omega/C$, where $C=\Binom{N}{\omega}$, is fixed from $D_\omega$ obtained directly from the decoder.}
\label{tab:fs-bitflip-new}
\begin{tabular}{cccc}
\hline
$d$ & $s_1$ & $s_2$ & $s_3$ \\
\hline
3 & 4.381 & 35.303 & -279.087 \\
5 & 7.323 & 145.077 & 511.673 \\
7 & 3.198 & 284.667 & 2428.236 \\
9 & 0.1213 & 280.228 & 1783.140 \\
\hline
\end{tabular}
\end{table}

\begin{table}[h]
\centering
\caption{\textbf{Superbranch fit coefficients $(s_1,s_2,s_3)$ for the depolarising channel} in \eqref{eqn: PS_surface_branch_series}. $f_{0}= D^{(dep)}_{\omega}/C$, where $C = \Binom{N}{\omega}3^{\omega}$, is fixed from $D^{(dep)}_\omega$ obtained from \cref{eqn: Aw_dep_mapping}.}
\label{tab:fs-depol-new}
\begin{tabular}{cccc}
\hline
$d$ & $s_1$ & $s_2$ & $s_3$ \\
\hline
3 & 3.778 & 21.486 & -27.545 \\
5 & 3.683 & 56.304 & 117.671 \\
7 & 1.294 & 51.561 & -241.690 \\
9 & 0.283 & 12.624 & -37.935 \\
\hline
\end{tabular}
\end{table}

\section{Estimators and Stratified Sampling} \label{appendix: sec: strat_sampling}

In this appendix, we describe the stratified sampling used to reduce estimator variance in the simulations. Because component-wise logical error rates vary sharply with the weight of the errors, stratification achieves significant variance reductions compared to the sampling overhead of PEC explored in \cref{appendix: sec: operational_definition}.

As introduced by Chen~\cite{chen2025fasterprobabilisticerrorcancellation} and Dai et al. (\emph{in prep.})~\cite{Dai2025StratifiedQPD}, we use stratified sampling to minimize the variance. For the identity branch, we stratify by the weight of the error $k$. Let $q_{I}$ be the logical failure probability when there are errors on exactly $k$ data qubits, i.e. $q_I(k)=\Pr(\text{logical error}\mid\text{exactly }k\text{ flips on data})$,
estimated by drawing uniformly random $k$-subsets and decoding. Then
\begin{equation}
P^{(0)}_L(p)=\sum_{k=0}^{N}\underbrace{\binom{N}{k}p^k(1-p)^{N-k}}_{\pi_k(p)}\; q_I(k).
\label{eq:PI-mixture}
\end{equation}Note that $q_{I}(k) = D_{k}/\Binom{N}{k}$, where  $D_k$ is the number of weight-$k$ error patterns that lead to a logical error under the chosen decoder. 
Furthermore, $f_0 = D_\omega/\binom{N}{\omega}$ in \cref{eqn: fail_prob_super_branch_branch_cond} and \cref{eqn: PS_surface_branch_series} coincides numerically with $q_I(\omega)$ in the code-capacity model.
However, $f_0$ is used to denote the logical failure probability of the weight-$\omega$ error \emph{inserted} by PEC,
whereas $q_I(k)$ refers to logical failure probability conditioned on errors drawn from the physical noise channel.

Given $m_k$ samples in stratum $k$ with empirical fail-rate $\hat q_I(k)$, the estimated logical error rate of the identity branch and its sample variance are given by
\begin{equation}
\begin{aligned}
\widehat P^{(0)}_L&=\sum_k \pi_k\,\hat q_I(k) \\ 
\mathrm{Var}(\widehat P^{(0)}_L)&=\sum_k \pi_k^2\,\frac{\hat q_I(k)\big(1-\hat q_I(k)\big)}{m_k}.
\end{aligned}
\label{eq:PI-var}
\end{equation}

For the superbranch, we stratify by the weight of overlapping errors, $u$, and the weight of extra errors, $r$. Let $q_S(u,r)$ be the logical failure probability conditioned on $(u,r)$, i.e. $q_S(u,r)=\Pr(\text{logical error}\mid u\text{ overlaps, }r \text{ extra errors})$. Then
\begin{equation}
\begin{aligned}
P^{(\omega)}_L(p)=\sum_{u=0}^{\omega}\sum_{r=0}^{N-\omega}
&\underbrace{\binom{\omega}{u}p^u(1-p)^{\omega-u}}_{\text{overlaps}} \\
&\underbrace{\binom{N-\omega}{r}p^r(1-p)^{N-\omega-r}}_{\text{extras}}
\; q_S(u,r).
\end{aligned}
\label{eq:PS-mixture}
\end{equation}
With $m_{u,r}$ samples in the $(u,r)$-stratum, the estimated logical error rate of the superbranch and its sample variance is given by

\begin{equation}
\begin{aligned}
\widehat P^{(\omega)}_L&=\sum_{u,r} w_{u,r}\,\hat q_S(u,r), \\
\mathrm{Var}(\widehat P^{(\omega)}_L)&=\sum_{u,r} w_{u,r}^2\,\frac{\hat q_S(u,r)\big(1-\hat q_S(u,r)\big)}{m_{u,r}},
\end{aligned}
\end{equation}
where $w_{u,r}$ is the product of the two binomial weights in \cref{eq:PS-mixture}. 

By combining the results of the two branches, we estimate the logical error rate and obtain its sample variance as follows:

\begin{equation}
\begin{aligned}
\widehat P_L^{\mathrm{PEC}}&= (1+G)\widehat P^{(0)}_L + G\widehat P^{(\omega)}_L\\
\mathrm{Var}(\widehat P_L^{\mathrm{PEC}})&=(1+G)^2\mathrm{Var}(\widehat P^{(0)}_L)+G^2\mathrm{Var}(\widehat P^{(\omega)}_L).
\end{aligned}
\end{equation} Since we estimated $P^{(0)}_L$ and $P^{(\omega)}_L$ from disjoint experiments with separate shot budgets, the covariance becomes zero, i.e.
$\mathrm{Cov}(\widehat P^{(0)}_L,\widehat P^{(\omega)}_L)=0$.

In practice, we truncated $k$ and $r$ to modest maximum, e.g. $k_{max}=\omega+2$ ($k_{max} = \omega+3$ for the depolarising noise) and $r_{max} = 3$ for the rotated surface code with bit-flip errors. For the depolarising channel, we uniformly sampled $X$, $Y$, and $Z$ from each stratum. Here, we report the number of shots per strata, i.e. the number of shots for each weight-$k$ in the identity branch and the number of shots for a pair of overlapping and extra gates in superbranch: ($m_{k}$, $m_{(u,r)}$) = ($800$k, $320k$) for both bit-flip and depolarising noise across all $d$ and $p$.

\section{More Result Plots} \label{appendix: sec: more_result_plots}

\begin{figure*}[htbp]
    \centering
    \subfloat[\label{fig: rc_xerror_d_3}]{\includegraphics[width=0.495\linewidth ]{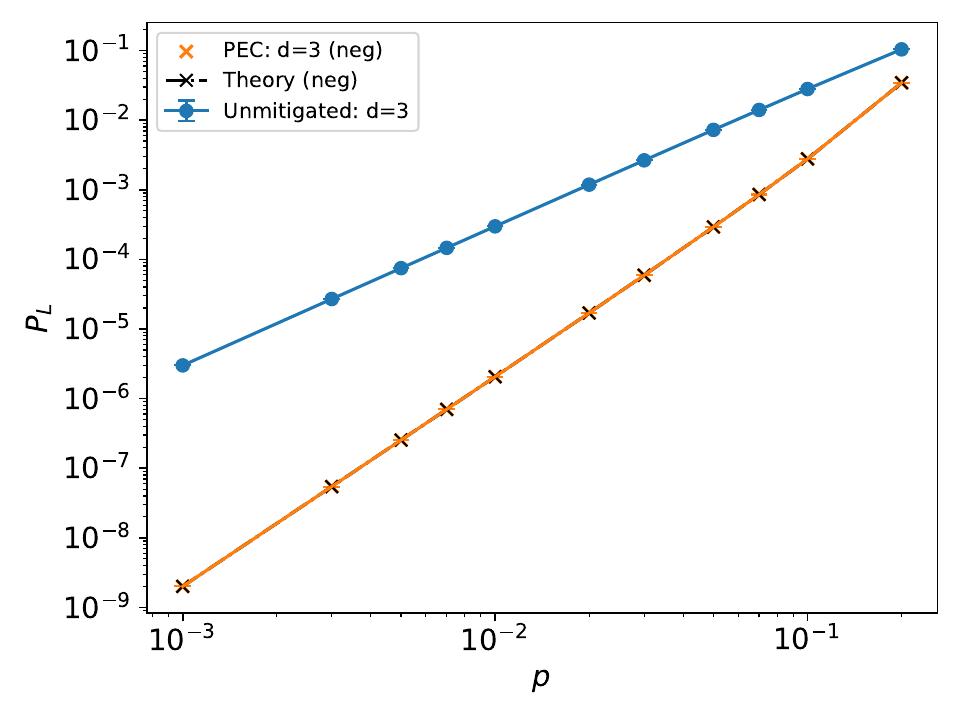}}~
    \subfloat[\label{fig: rc_xerror_d_5}]{\includegraphics[width=0.495\textwidth]{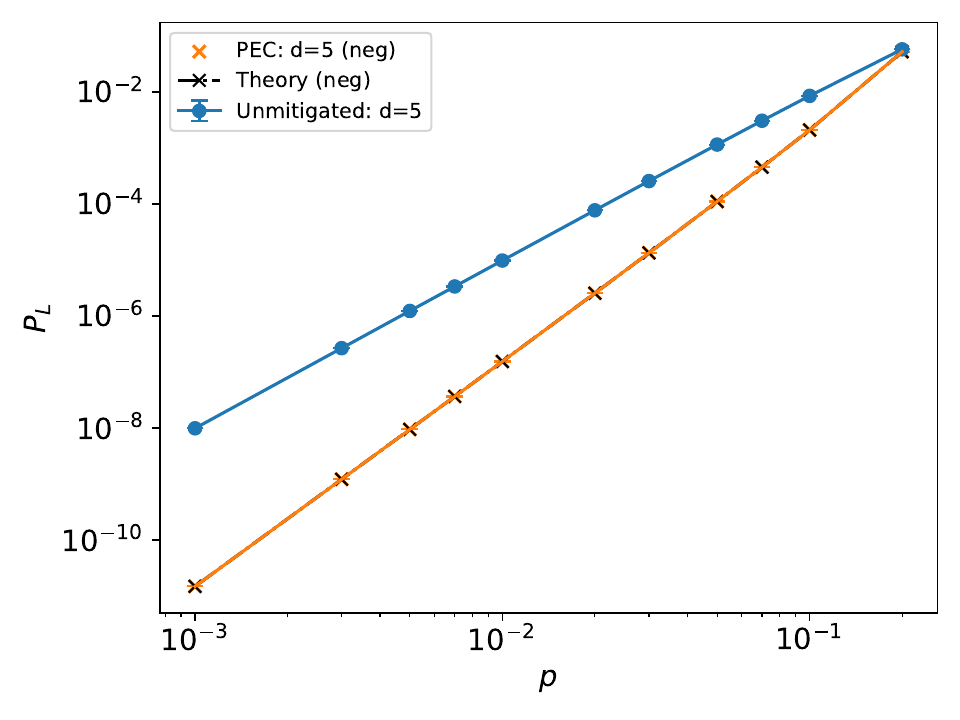}}
    \hfill
    \subfloat[\label{fig: rc_xerror_d_7}]{\includegraphics[width=0.495\textwidth]{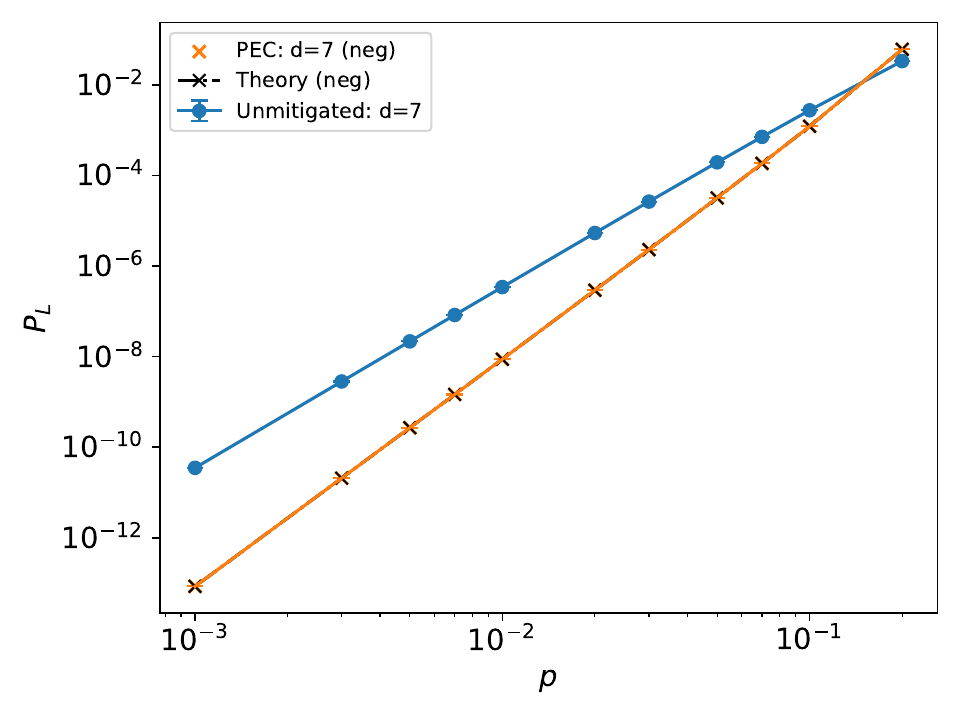}}~
    \subfloat[\label{fig: rc_xerror_d_9}]{\includegraphics[width=0.495\textwidth]{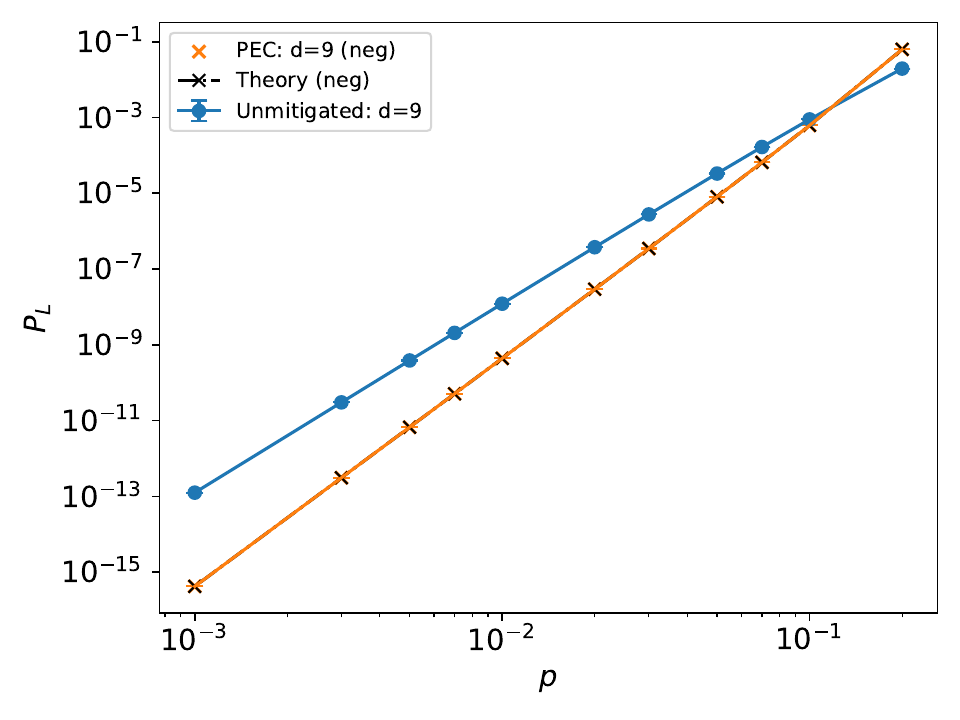}}

    \caption{The results of the memory experiment of the repetition code under the code-capacity model with bit-flip errors: (blue) Unmitigated repetition code and (orange) repetition code with PEC (black; dotted) theoretical predictions by \cref{appendix: subsec: branch_cond_rep_code}. Note that the standard deviation of both repetition code with and without PEC is zero due to stratification, i.e. the decoder outcome is deterministic for the given stratum. The resulting slope of the linear fitting for the repetition code without PEC and with PEC are as follows: (a) $d=3$: $1.98$ (No PEC), $3.11$ (PEC), (b) $d=5$: $2.95$ (No PEC), $4.11$ (PEC), (c) $d=7$: $3.92$ (No PEC), $5.12$ (PEC), and (d) $d=9$: $4.89$ (No PEC), $6.13$(PEC).}
\label{fig: rc_xerror}
\end{figure*}

\begin{figure*}[htbp]
    \centering
    \subfloat[\label{fig: sc_xerror_d_3}]{\includegraphics[width=0.495\linewidth ]{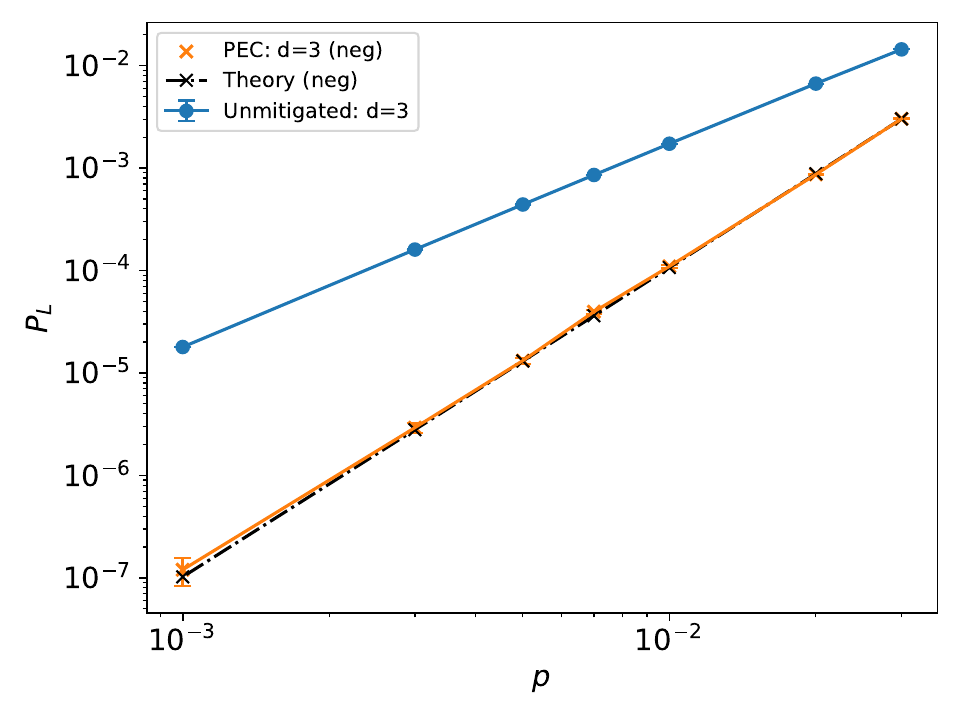}}~
    \subfloat[\label{fig: sc_xerror_d_5}]{\includegraphics[width=0.495\textwidth]{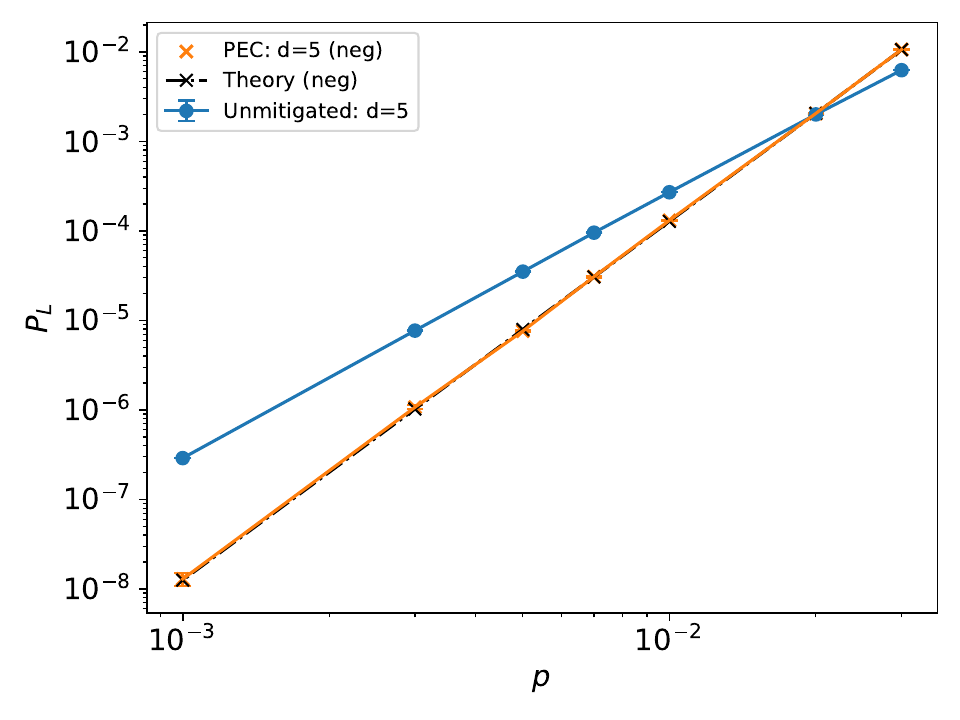}}
    \hfill
    \subfloat[\label{fig: sc_xerror_d_7}]{\includegraphics[width=0.495\textwidth]{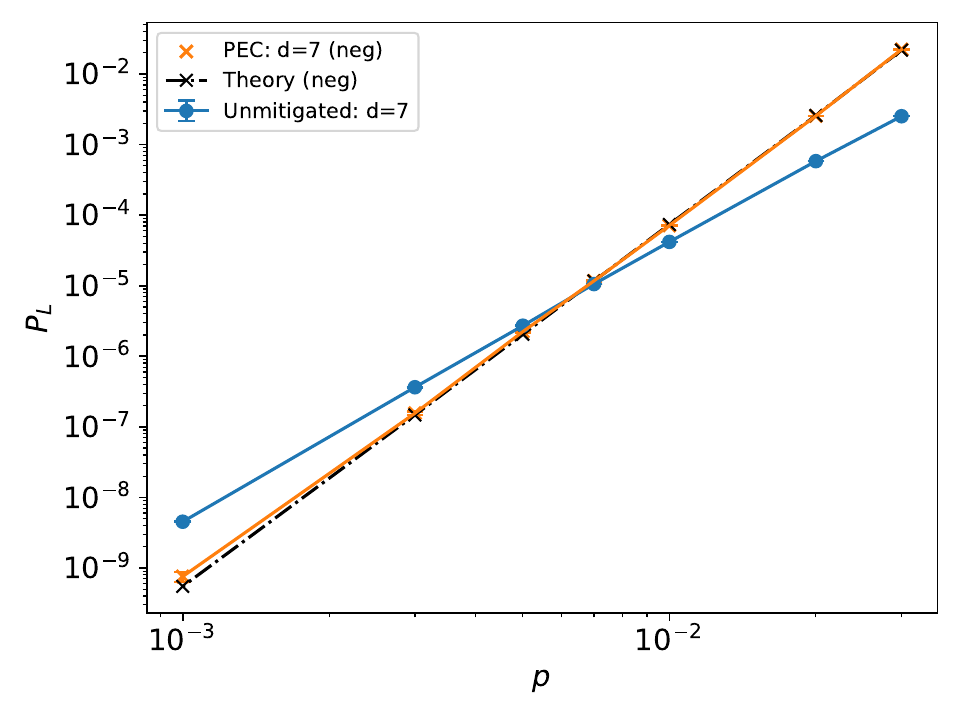}}~
    \subfloat[\label{fig: sc_xerror_d_9}]{\includegraphics[width=0.495\textwidth]{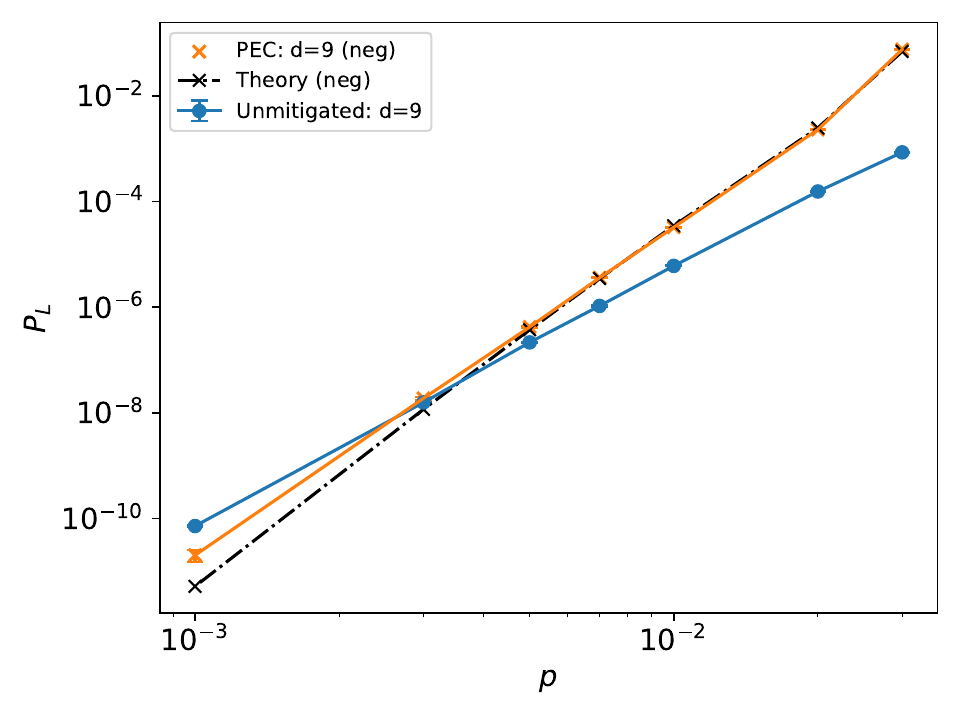}}

    \caption{The results of the memory experiment of the rotated surface code under the code-capacity model with bit-flip errors: (blue) Unmitigated rotated surface code and (orange) rotated surface code with PEC (black; dotted) theoretical predictions by \cref{appendix: subsec: branch_cond_surface_code}. Note that the standard deviations are not zero as the decoding is not deterministic given fixed weights, $k$, ( or overlapping and extra weights, $(u, r)$, for the superbranch). The resulting slope of the linear fitting for the rotated surface code without PEC and with PEC are as follows: (a) $d=3$: $1.97$ (No PEC), $2.98$ (PEC), (b) $d=5$: $2.94$ (No PEC), $4.00$ (PEC), (c) $d=7$: $3.90$ (No PEC), $5.06$ (PEC), and (d) $d=9$: $4.81$ (No PEC), $6.37$ (PEC). Note that, for the PEC results, there are more contributions from the higher order $p$ terms for the increasing distance. For example, for $d=9$, the curve bend upwards at $p=3\times10^{-2}$, resulting in slope of $\lceil\frac{9}{2}\rceil+1 =6 \leq 6.37$}
\label{fig: sc_xerror}
\end{figure*}

\begin{figure*}[htbp]
    \centering
    \subfloat[\label{fig: sc_depol1_error_d_3}]{\includegraphics[width=0.495\linewidth ]{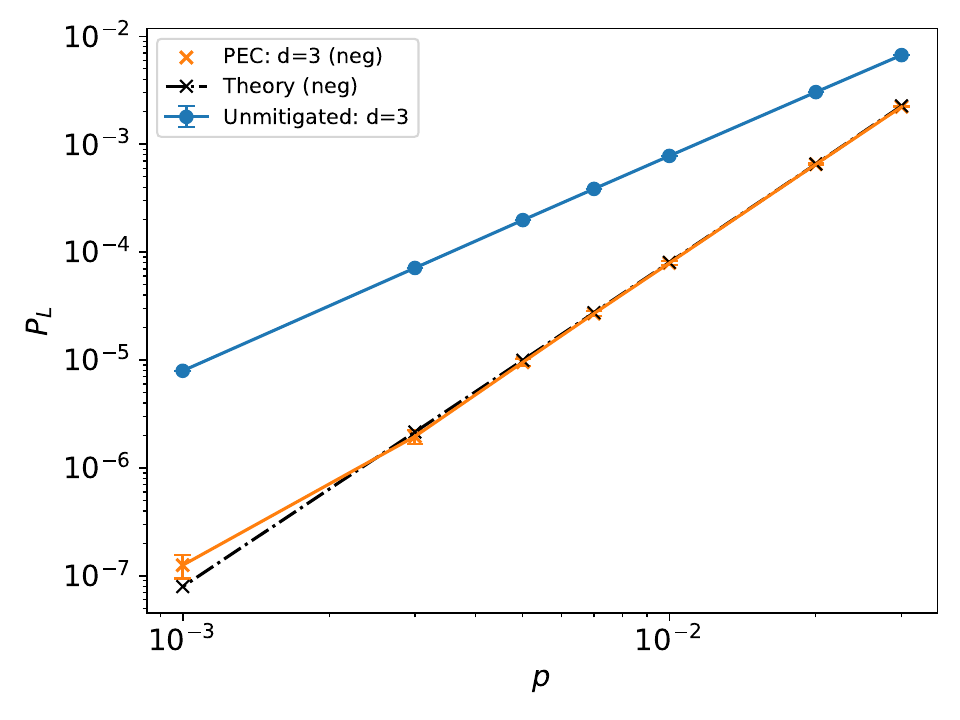}}~
    \subfloat[\label{fig: sc_depol1_error_d_5}]{\includegraphics[width=0.495\textwidth]{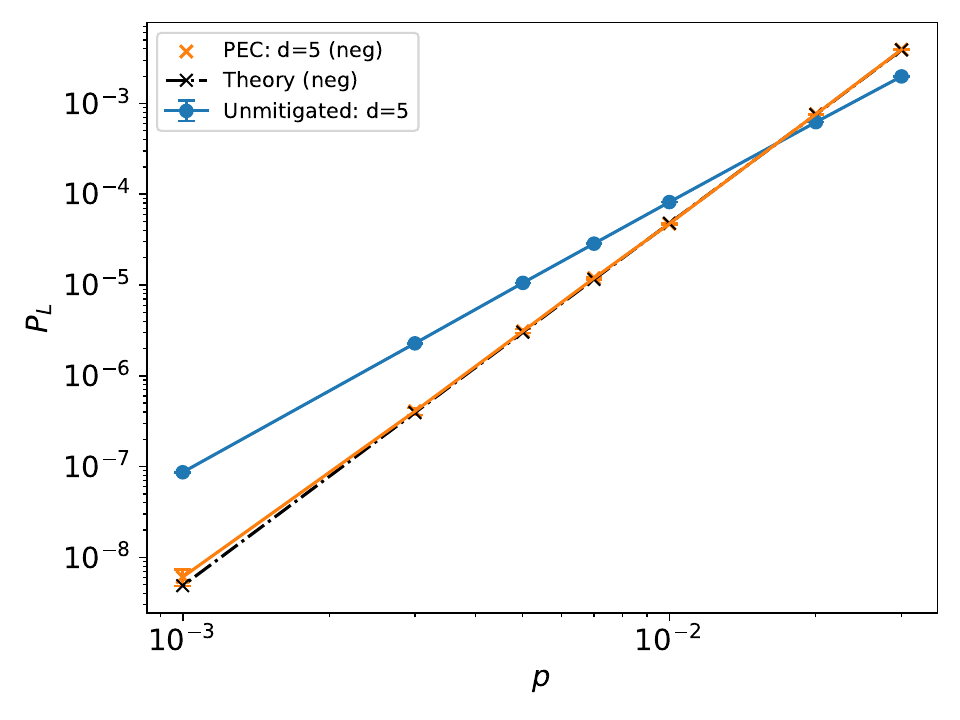}}
    \hfill
    \subfloat[\label{fig: sc_depol1_error_d_7}]{\includegraphics[width=0.495\textwidth]{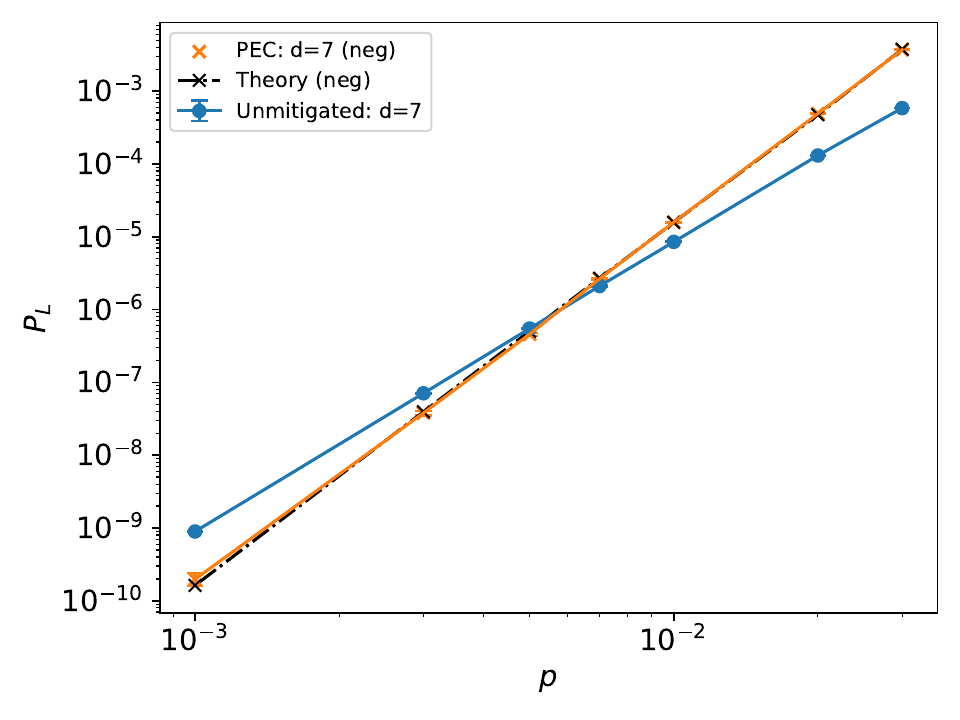}}~
    \subfloat[\label{fig: sc_depol1_error_d_9}]{\includegraphics[width=0.495\textwidth]{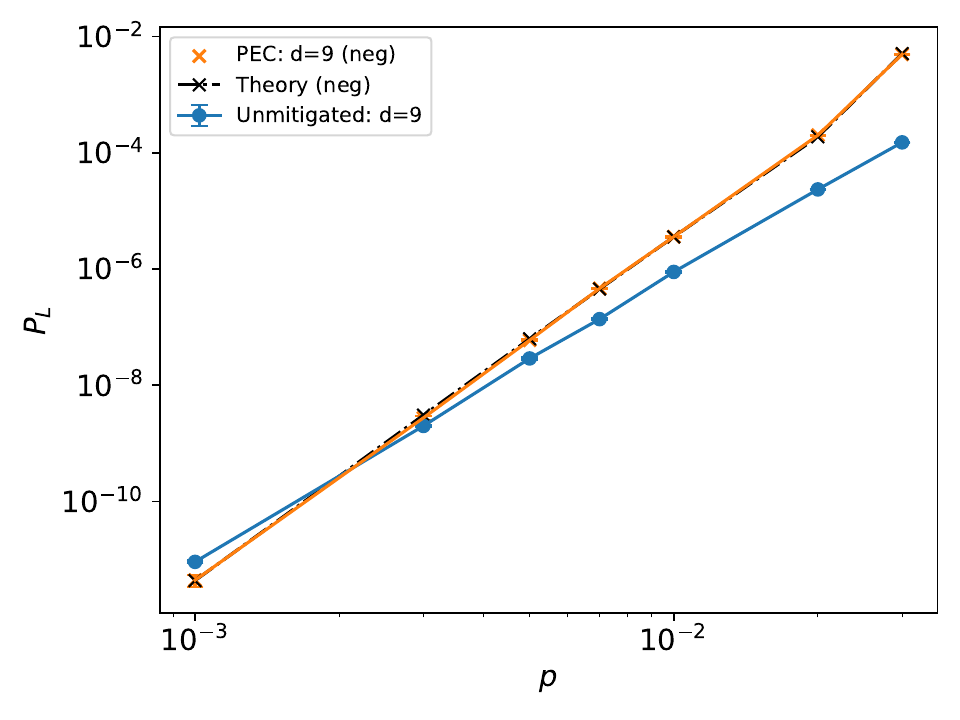}}

    \caption{The results of the memory experiment of the rotated surface code under the code-capacity model with depolarising noise: (blue) Unmitigated rotated surface code and (orange) rotated surface code with PEC (black; dotted) theoretical predictions by \cref{appendix: subsec: branch_cond_surface_code}. Note that the standard deviations are not zero as the decoding is not deterministic given fixed weights, $k$, ( or overlapping and extra weights, $(u, r)$, for the superbranch). The resulting slope of the linear fitting for the rotated surface code without PEC and with PEC are as follows: (a) $d=3$: $1.98$ (No PEC), $2.91$ (PEC), (b) $d=5$: $2.96$ (No PEC), $3.94$ (PEC), (c) $d=7$: $3.95$ (No PEC), $4.94$ (PEC), and (d) $d=9$: $4.90$ (No PEC), $6.04$ (PEC). Note that, for the PEC results, there are more contributions from the higher order $p$ terms for the increasing distance. For example, for $d=9$, the curve bend upwards at $p=3\times10^{-2}$, resulting in slope of $\lceil\frac{9}{2}\rceil+1 =6 \leq 6.04$}
\label{fig: sc_depol1_error}
\end{figure*}

In this appendix, we provide a detailed comparison between numerical simulations and theoretical predictions for the logical error rates with and without PEC. The theoretical curves (black dotted lines) are obtained using the analytic and semi-analytic expressions of component-wise logical error rates developed in \cref{appendix: sec: branch_cond_rep_code_surface_code}. The numerical simulations were performed by stratified sampling in \cref{appendix: sec: strat_sampling}. Across all code distances and noise models, the theory reproduces the full numerical logical error curves with high accuracy. The close agreement with numerical data confirms that PEC on physical qubits before QEC can effectively mitigate the leading order in the logical error rate. See \cref{fig: rc_xerror}, \cref{fig: sc_xerror}, and \cref{fig: sc_depol1_error}.

\begin{figure*}
	\centering
	\subfloat[]{
    \includegraphics[width=0.495\linewidth]{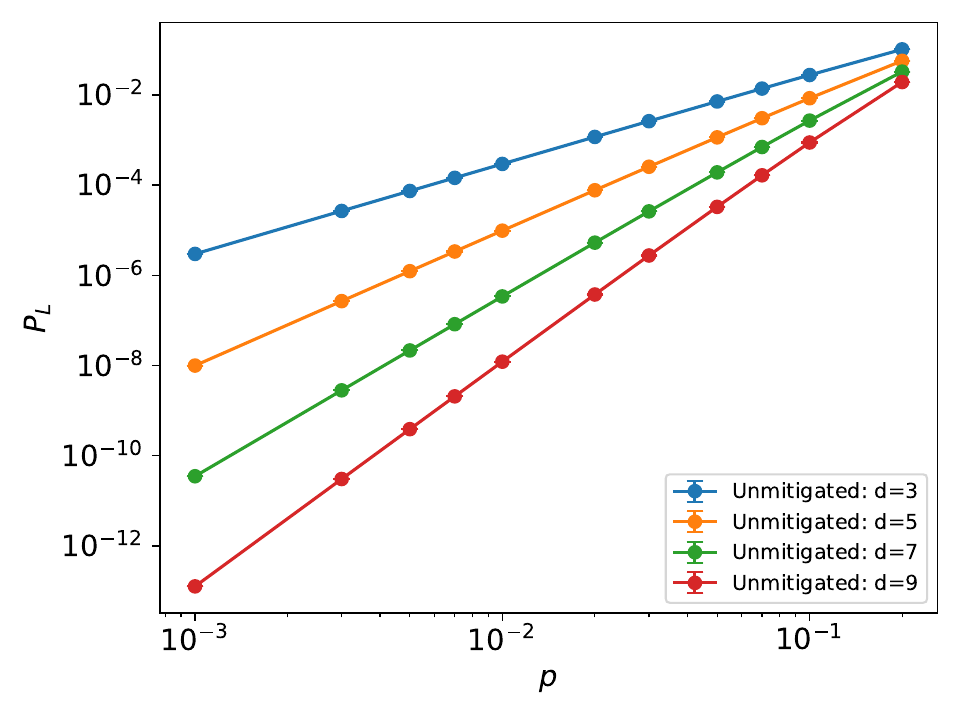}
    \label{fig: rc_x_error_vanilla_threshold}
    }
    \subfloat[]{
   \includegraphics[width=0.495\linewidth]{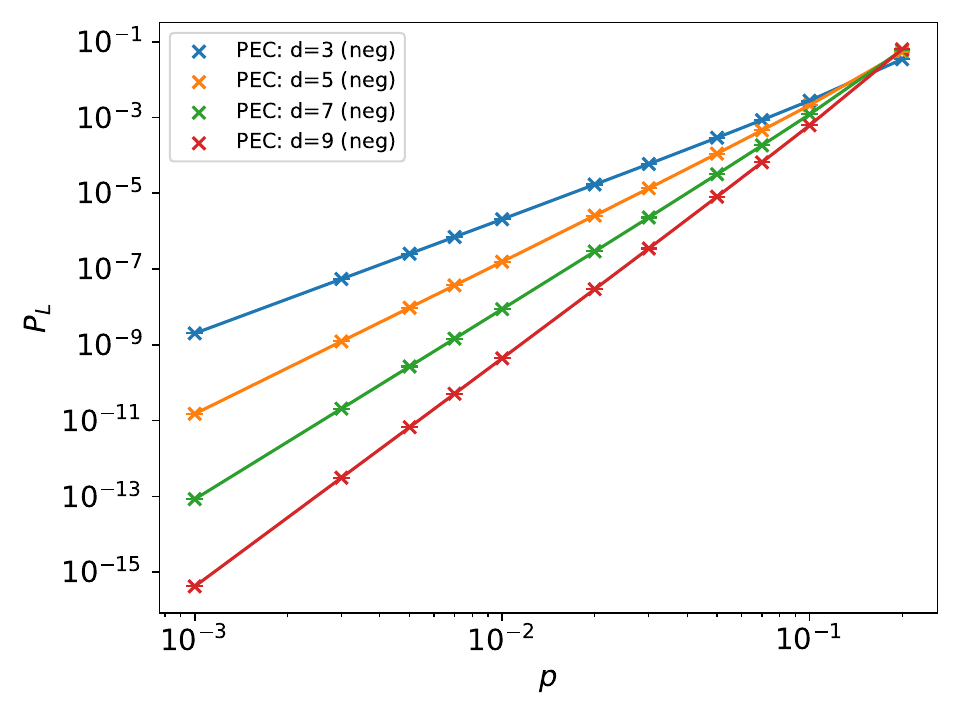}
   \label{fig: rc_x_error_pec_threshold} 
      }
    \caption{Threshold plots for (a) Unmitigated and (b) PEC repetition code, with thresholds of $0.5$~\cite{Vodola_2022} and $0.19$, respectively. The threshold of PEC was estimated by the intersection of the two largest distances, i.e. $d=7, \: 9$. Due to the steeper slopes, the repetition code with PEC has lower threshold than the unmitigated repetition code.}
    \label{fig: rc_x_error_threshold}
\end{figure*}

\begin{figure*}
	\centering
	\subfloat[]{
    \includegraphics[width=0.495\linewidth]{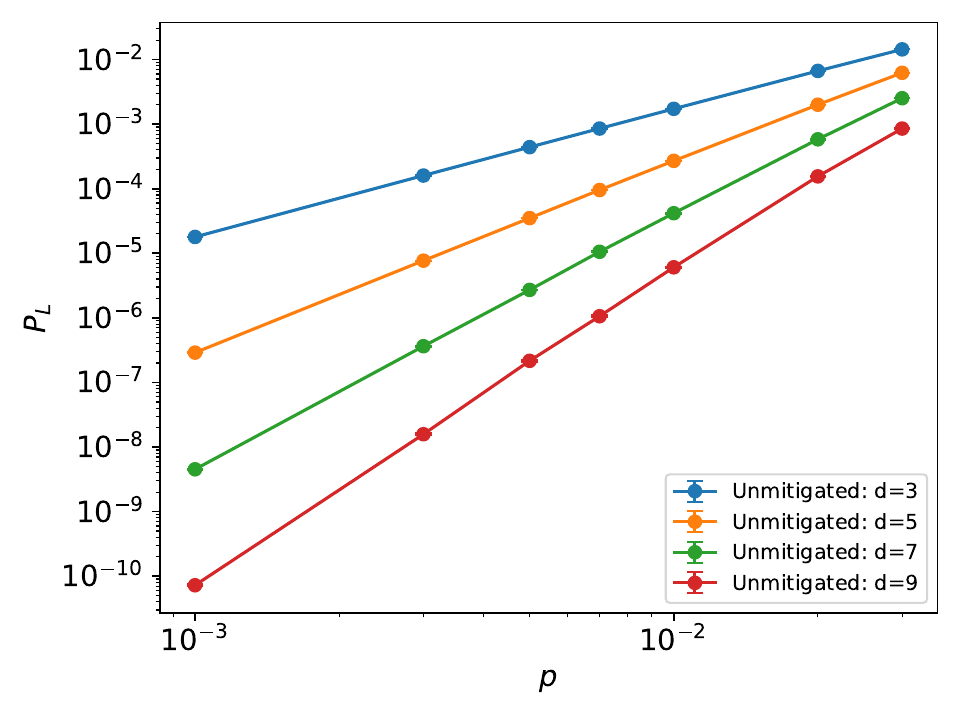}
    \label{fig: sc_x_error_vanilla_threshold}
    }
    \subfloat[]{
   \includegraphics[width=0.495\linewidth]{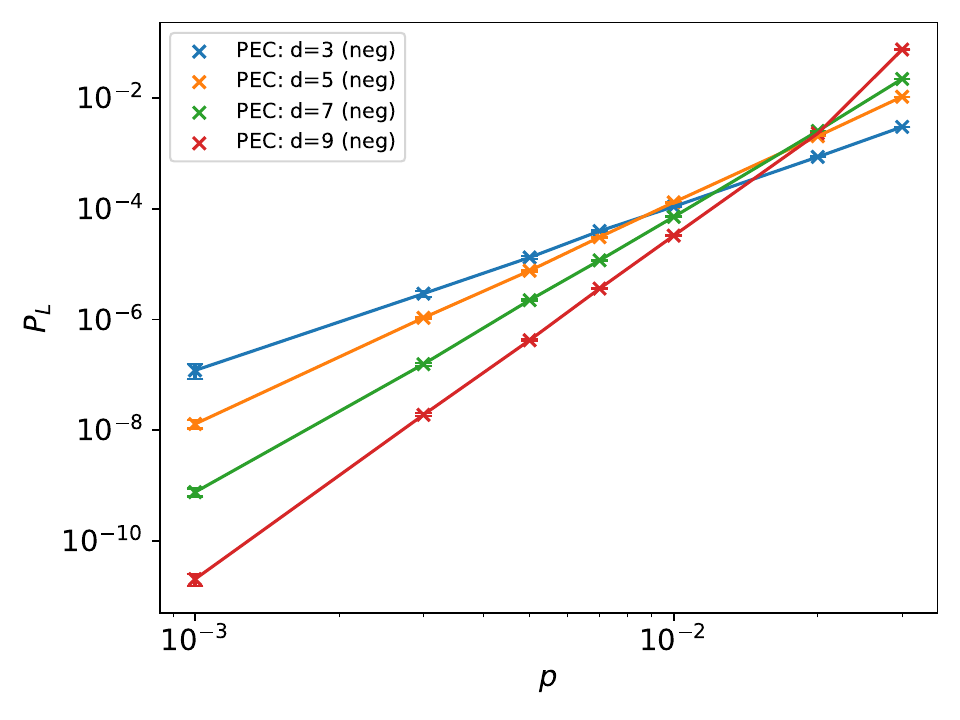}
   \label{fig: sc_x_error_pec_threshold} 
      }
    \caption{Threshold plots for (a) Unmitigated and (b) PEC surface code under bit-flip noise, with estimated thresholds of $0.1094$~\cite{Honecker_2001, Dennis_2002} and $0.023$, respectively. The threshold of PEC was estimated by the intersection of the two largest distances, i.e. $d=7, \: 9$. Due to the steeper slopes, the surface code with PEC has lower threshold than the unmitigated surface code.}
    \label{fig: sc_x_error_threshold}
\end{figure*}

\begin{figure*}
	\centering
	\subfloat[]{
    \includegraphics[width=0.495\linewidth]{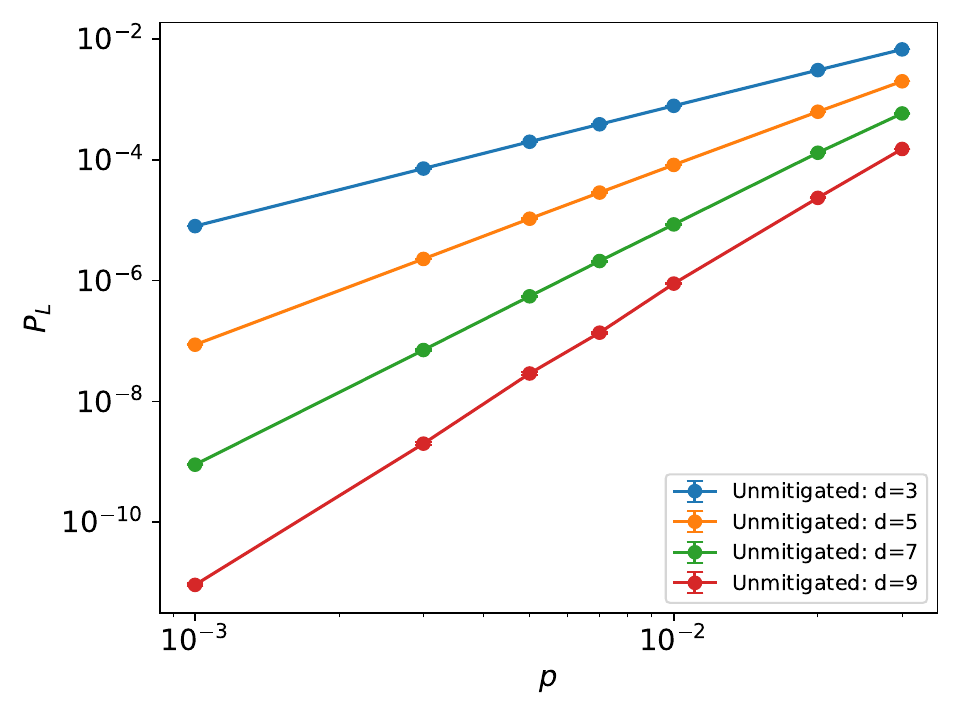}
    \label{fig: sc_depol1_error_vanilla_threshold}
    }
    \subfloat[]{
   \includegraphics[width=0.495\linewidth]{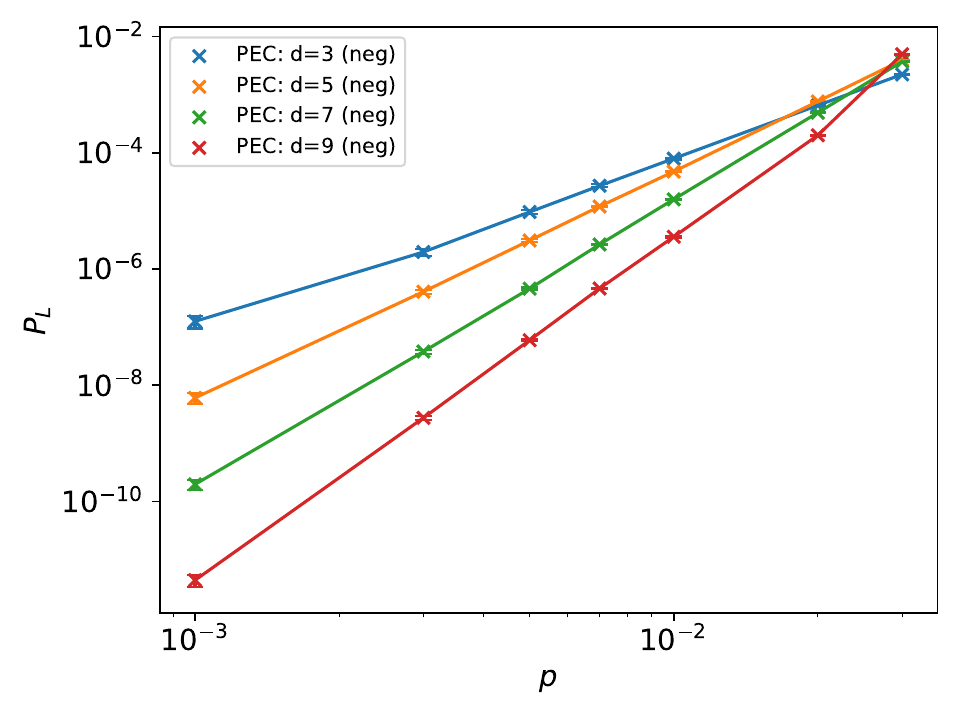}
   \label{fig: sc_depol1_error_pec_threshold} 
      }
    \caption{Threshold plots for (a) Unmitigated and (b) PEC surface code under depolarising noise, with the estimated thresholds of $0.145$~\cite{Kuo_2022} and $0.029$, respectively. The threshold of PEC was estimated by the intersection of the two largest distances, i.e. $d=7, \: 9$. Due to the steeper slopes, the surface code with PEC has lower threshold than the unmitigated surface code.}
    \label{fig: sc_depol1_error_threshold}
\end{figure*}

\cref{fig: rc_x_error_threshold}, \cref{fig: sc_x_error_threshold}, and \cref{fig: sc_depol1_error_threshold} are the threshold plots of repetition code under bit-flip noise, surface code under bit-flip noise, and surface code under depolarising noise, respectively. Because PEC lifts the logical error rate by one order to $\order{p^{\omega+1}}$, the logical-error curves steepen and therefore the threshold decreases $(p_{\mathrm{th}}^{\text{PEC}}<p_{\mathrm{th}}^{\text{Unmitigated}})$. Note that \cref{fig: sc_x_error_threshold} and \cref{fig: sc_depol1_error_threshold} reproduce the same data as \cref{fig:P_L_plots}, but with the unmitigated and PEC logical error rates shown in separate panels to more clearly illustrate the threshold behaviour.

\bibliography{refs}

@article{gidney2021stim,
  doi = {10.22331/q-2021-07-06-497},
  url = {https://doi.org/10.22331/q-2021-07-06-497},
  title = {Stim: a fast stabilizer circuit simulator},
  author = {Gidney, Craig},
  journal = {{Quantum}},
  issn = {2521-327X},
  publisher = {{Verein zur F{\"{o}}rderung des Open Access Publizierens
                in den Quantenwissenschaften}},
  volume = 5,
  pages = 497,
  month = jul,
  year = 2021
}

@article{Higgott2025sparseblossom,
  doi = {10.22331/q-2025-01-20-1600},
  url = {https://doi.org/10.22331/q-2025-01-20-1600},
  title = {Sparse {B}lossom: correcting a million errors per core second with minimum-weight matching},
  author = {Higgott, Oscar and Gidney, Craig},
  journal = {{Quantum}},
  issn = {2521-327X},
  publisher = {{Verein zur F{\"{o}}rderung des Open Access Publizierens in den Quantenwissenschaften}},
  volume = {9},
  pages = {1600},
  month = jan,
  year = {2025}
}

@misc{zhang2025demonstratingquantumerrormitigation,
      title={Demonstrating quantum error mitigation on logical qubits}, 
      author={Aosai Zhang and Haipeng Xie and Yu Gao and Jia-Nan Yang and Zehang Bao and Zitian Zhu and Jiachen Chen and Ning Wang and Chuanyu Zhang and Jiarun Zhong and Shibo Xu and Ke Wang and Yaozu Wu and Feitong Jin and Xuhao Zhu and Yiren Zou and Ziqi Tan and Zhengyi Cui and Fanhao Shen and Tingting Li and Yihang Han and Yiyang He and Gongyu Liu and Jiayuan Shen and Han Wang and Yanzhe Wang and Hang Dong and Jinfeng Deng and Hekang Li and Zhen Wang and Chao Song and Qiujiang Guo and Pengfei Zhang and Ying Li and H. Wang},
      year={2025},
      eprint={2501.09079},
      archivePrefix={arXiv},
      primaryClass={quant-ph},
      url={https://arxiv.org/abs/2501.09079}, 
}

@article{Cai_2023,
   title={Quantum error mitigation},
   volume={95},
   ISSN={1539-0756},
   url={http://dx.doi.org/10.1103/RevModPhys.95.045005},
   DOI={10.1103/revmodphys.95.045005},
   number={4},
   journal={Reviews of Modern Physics},
   publisher={American Physical Society (APS)},
   author={Cai, Zhenyu and Babbush, Ryan and Benjamin, Simon C. and Endo, Suguru and Huggins, William J. and Li, Ying and McClean, Jarrod R. and O’Brien, Thomas E.},
   year={2023},
   month=dec }

@article{Terhal_2015,
   title={Quantum error correction for quantum memories},
   volume={87},
   ISSN={1539-0756},
   url={http://dx.doi.org/10.1103/RevModPhys.87.307},
   DOI={10.1103/revmodphys.87.307},
   number={2},
   journal={Reviews of Modern Physics},
   publisher={American Physical Society (APS)},
   author={Terhal, Barbara M.},
   year={2015},
   month=apr, pages={307–346} }

@article{Fowler_2012,
   title={Surface codes: Towards practical large-scale quantum computation},
   volume={86},
   ISSN={1094-1622},
   url={http://dx.doi.org/10.1103/PhysRevA.86.032324},
   DOI={10.1103/physreva.86.032324},
   number={3},
   journal={Physical Review A},
   publisher={American Physical Society (APS)},
   author={Fowler, Austin G. and Mariantoni, Matteo and Martinis, John M. and Cleland, Andrew N.},
   year={2012},
   month=sep }

@book{gottesman1997stabilizer,
  title={Stabilizer codes and quantum error correction},
  author={Gottesman, Daniel},
  year={1997},
  publisher={California Institute of Technology}
}

@article{Endo_2018,
   title={Practical Quantum Error Mitigation for Near-Future Applications},
   volume={8},
   ISSN={2160-3308},
   url={http://dx.doi.org/10.1103/PhysRevX.8.031027},
   DOI={10.1103/physrevx.8.031027},
   number={3},
   journal={Physical Review X},
   publisher={American Physical Society (APS)},
   author={Endo, Suguru and Benjamin, Simon C. and Li, Ying},
   year={2018},
   month=jul }

@article{suzuki2022,
  title = {Quantum Error Mitigation as a Universal Error Reduction Technique: Applications from the NISQ to the Fault-Tolerant Quantum Computing Eras},
  author = {Suzuki, Yasunari and Endo, Suguru and Fujii, Keisuke and Tokunaga, Yuuki},
  journal = {PRX Quantum},
  volume = {3},
  issue = {1},
  pages = {010345},
  numpages = {33},
  year = {2022},
  month = {Mar},
  publisher = {American Physical Society},
  doi = {10.1103/PRXQuantum.3.010345},
  url = {https://link.aps.org/doi/10.1103/PRXQuantum.3.010345}
}

@article{kandala2019error,
  title={Error mitigation extends the computational reach of a noisy quantum processor},
  author={Kandala, Abhinav and Temme, Kristan and C{\'o}rcoles, Antonio D and Mezzacapo, Antonio and Chow, Jerry M and Gambetta, Jay M},
  journal={Nature},
  volume={567},
  number={7749},
  pages={491--495},
  year={2019},
  publisher={Nature Publishing Group UK London}
}

@article{google2023suppressing,
   title={Suppressing quantum errors by scaling a surface code logical qubit},
   volume={614},
   ISSN={1476-4687},
   url={http://dx.doi.org/10.1038/s41586-022-05434-1},
   DOI={10.1038/s41586-022-05434-1},
   number={7949},
   journal={Nature},
   publisher={Springer Science and Business Media LLC},
   author={Acharya, Rajeev and Aleiner, Igor and Allen, Richard and Andersen, Trond I. and Ansmann, Markus and Arute, Frank and Arya, Kunal and Asfaw, Abraham and Atalaya, Juan and Babbush, Ryan and Bacon, Dave and Bardin, Joseph C. and Basso, Joao and Bengtsson, Andreas and Boixo, Sergio and Bortoli, Gina and Bourassa, Alexandre and Bovaird, Jenna and Brill, Leon and Broughton, Michael and Buckley, Bob B. and Buell, David A. and Burger, Tim and Burkett, Brian and Bushnell, Nicholas and Chen, Yu and Chen, Zijun and Chiaro, Ben and Cogan, Josh and Collins, Roberto and Conner, Paul and Courtney, William and Crook, Alexander L. and Curtin, Ben and Debroy, Dripto M. and Del Toro Barba, Alexander and Demura, Sean and Dunsworth, Andrew and Eppens, Daniel and Erickson, Catherine and Faoro, Lara and Farhi, Edward and Fatemi, Reza and Flores Burgos, Leslie and Forati, Ebrahim and Fowler, Austin G. and Foxen, Brooks and Giang, William and Gidney, Craig and Gilboa, Dar and Giustina, Marissa and Grajales Dau, Alejandro and Gross, Jonathan A. and Habegger, Steve and Hamilton, Michael C. and Harrigan, Matthew P. and Harrington, Sean D. and Higgott, Oscar and Hilton, Jeremy and Hoffmann, Markus and Hong, Sabrina and Huang, Trent and Huff, Ashley and Huggins, William J. and Ioffe, Lev B. and Isakov, Sergei V. and Iveland, Justin and Jeffrey, Evan and Jiang, Zhang and Jones, Cody and Juhas, Pavol and Kafri, Dvir and Kechedzhi, Kostyantyn and Kelly, Julian and Khattar, Tanuj and Khezri, Mostafa and Kieferová, Mária and Kim, Seon and Kitaev, Alexei and Klimov, Paul V. and Klots, Andrey R. and Korotkov, Alexander N. and Kostritsa, Fedor and Kreikebaum, John Mark and Landhuis, David and Laptev, Pavel and Lau, Kim-Ming and Laws, Lily and Lee, Joonho and Lee, Kenny and Lester, Brian J. and Lill, Alexander and Liu, Wayne and Locharla, Aditya and Lucero, Erik and Malone, Fionn D. and Marshall, Jeffrey and Martin, Orion and McClean, Jarrod R. and McCourt, Trevor and McEwen, Matt and Megrant, Anthony and Meurer Costa, Bernardo and Mi, Xiao and Miao, Kevin C. and Mohseni, Masoud and Montazeri, Shirin and Morvan, Alexis and Mount, Emily and Mruczkiewicz, Wojciech and Naaman, Ofer and Neeley, Matthew and Neill, Charles and Nersisyan, Ani and Neven, Hartmut and Newman, Michael and Ng, Jiun How and Nguyen, Anthony and Nguyen, Murray and Niu, Murphy Yuezhen and O’Brien, Thomas E. and Opremcak, Alex and Platt, John and Petukhov, Andre and Potter, Rebecca and Pryadko, Leonid P. and Quintana, Chris and Roushan, Pedram and Rubin, Nicholas C. and Saei, Negar and Sank, Daniel and Sankaragomathi, Kannan and Satzinger, Kevin J. and Schurkus, Henry F. and Schuster, Christopher and Shearn, Michael J. and Shorter, Aaron and Shvarts, Vladimir and Skruzny, Jindra and Smelyanskiy, Vadim and Smith, W. Clarke and Sterling, George and Strain, Doug and Szalay, Marco and Torres, Alfredo and Vidal, Guifre and Villalonga, Benjamin and Vollgraff Heidweiller, Catherine and White, Theodore and Xing, Cheng and Yao, Z. Jamie and Yeh, Ping and Yoo, Juhwan and Young, Grayson and Zalcman, Adam and Zhang, Yaxing and Zhu, Ningfeng},
   year={2023},
   month=feb, pages={676–681} }

@article{krinner2022realizing,
  title={Realizing repeated quantum error correction in a distance-three surface code},
  author={Krinner, Sebastian and Lacroix, Nathan and Remm, Ants and Di Paolo, Agustin and Genois, Elie and Leroux, Catherine and Hellings, Christoph and Lazar, Stefania and Swiadek, Francois and Herrmann, Johannes and others},
  journal={Nature},
  volume={605},
  number={7911},
  pages={669--674},
  year={2022},
  publisher={Nature Publishing Group UK London}
}

@article{google2025quantum,
   title={Quantum error correction below the surface code threshold},
   volume={638},
   ISSN={1476-4687},
   url={http://dx.doi.org/10.1038/s41586-024-08449-y},
   DOI={10.1038/s41586-024-08449-y},
   number={8052},
   journal={Nature},
   publisher={Springer Science and Business Media LLC},
   author={Acharya, Rajeev and Abanin, Dmitry A. and Aghababaie-Beni, Laleh and Aleiner, Igor and Andersen, Trond I. and Ansmann, Markus and Arute, Frank and Arya, Kunal and Asfaw, Abraham and Astrakhantsev, Nikita and Atalaya, Juan and Babbush, Ryan and Bacon, Dave and Ballard, Brian and Bardin, Joseph C. and Bausch, Johannes and Bengtsson, Andreas and Bilmes, Alexander and Blackwell, Sam and Boixo, Sergio and Bortoli, Gina and Bourassa, Alexandre and Bovaird, Jenna and Brill, Leon and Broughton, Michael and Browne, David A. and Buchea, Brett and Buckley, Bob B. and Buell, David A. and Burger, Tim and Burkett, Brian and Bushnell, Nicholas and Cabrera, Anthony and Campero, Juan and Chang, Hung-Shen and Chen, Yu and Chen, Zijun and Chiaro, Ben and Chik, Desmond and Chou, Charina and Claes, Jahan and Cleland, Agnetta Y. and Cogan, Josh and Collins, Roberto and Conner, Paul and Courtney, William and Crook, Alexander L. and Curtin, Ben and Das, Sayan and Davies, Alex and De Lorenzo, Laura and Debroy, Dripto M. and Demura, Sean and Devoret, Michel and Di Paolo, Agustin and Donohoe, Paul and Drozdov, Ilya and Dunsworth, Andrew and Earle, Clint and Edlich, Thomas and Eickbusch, Alec and Elbag, Aviv Moshe and Elzouka, Mahmoud and Erickson, Catherine and Faoro, Lara and Farhi, Edward and Ferreira, Vinicius S. and Burgos, Leslie Flores and Forati, Ebrahim and Fowler, Austin G. and Foxen, Brooks and Ganjam, Suhas and Garcia, Gonzalo and Gasca, Robert and Genois, \'{E}lie and Giang, William and Gidney, Craig and Gilboa, Dar and Gosula, Raja and Dau, Alejandro Grajales and Graumann, Dietrich and Greene, Alex and Gross, Jonathan A. and Habegger, Steve and Hall, John and Hamilton, Michael C. and Hansen, Monica and Harrigan, Matthew P. and Harrington, Sean D. and Heras, Francisco J. H. and Heslin, Stephen and Heu, Paula and Higgott, Oscar and Hill, Gordon and Hilton, Jeremy and Holland, George and Hong, Sabrina and Huang, Hsin-Yuan and Huff, Ashley and Huggins, William J. and Ioffe, Lev B. and Isakov, Sergei V. and Iveland, Justin and Jeffrey, Evan and Jiang, Zhang and Jones, Cody and Jordan, Stephen and Joshi, Chaitali and Juhas, Pavol and Kafri, Dvir and Kang, Hui and Karamlou, Amir H. and Kechedzhi, Kostyantyn and Kelly, Julian and Khaire, Trupti and Khattar, Tanuj and Khezri, Mostafa and Kim, Seon and Klimov, Paul V. and Klots, Andrey R. and Kobrin, Bryce and Kohli, Pushmeet and Korotkov, Alexander N. and Kostritsa, Fedor and Kothari, Robin and Kozlovskii, Borislav and Kreikebaum, John Mark and Kurilovich, Vladislav D. and Lacroix, Nathan and Landhuis, David and Lange-Dei, Tiano and Langley, Brandon W. and Laptev, Pavel and Lau, Kim-Ming and Le Guevel, Loïck and Ledford, Justin and Lee, Joonho and Lee, Kenny and Lensky, Yuri D. and Leon, Shannon and Lester, Brian J. and Li, Wing Yan and Li, Yin and Lill, Alexander T. and Liu, Wayne and Livingston, William P. and Locharla, Aditya and Lucero, Erik and Lundahl, Daniel and Lunt, Aaron and Madhuk, Sid and Malone, Fionn D. and Maloney, Ashley and Mandrà, Salvatore and Manyika, James and Martin, Leigh S. and Martin, Orion and Martin, Steven and Maxfield, Cameron and McClean, Jarrod R. and McEwen, Matt and Meeks, Seneca and Megrant, Anthony and Mi, Xiao and Miao, Kevin C. and Mieszala, Amanda and Molavi, Reza and Molina, Sebastian and Montazeri, Shirin and Morvan, Alexis and Movassagh, Ramis and Mruczkiewicz, Wojciech and Naaman, Ofer and Neeley, Matthew and Neill, Charles and Nersisyan, Ani and Neven, Hartmut and Newman, Michael and Ng, Jiun How and Nguyen, Anthony and Nguyen, Murray and Ni, Chia-Hung and Niu, Murphy Yuezhen and O’Brien, Thomas E. and Oliver, William D. and Opremcak, Alex and Ottosson, Kristoffer and Petukhov, Andre and Pizzuto, Alex and Platt, John and Potter, Rebecca and Pritchard, Orion and Pryadko, Leonid P. and Quintana, Chris and Ramachandran, Ganesh and Reagor, Matthew J. and Redding, John and Rhodes, David M. and Roberts, Gabrielle and Rosenberg, Eliott and Rosenfeld, Emma and Roushan, Pedram and Rubin, Nicholas C. and Saei, Negar and Sank, Daniel and Sankaragomathi, Kannan and Satzinger, Kevin J. and Schurkus, Henry F. and Schuster, Christopher and Senior, Andrew W. and Shearn, Michael J. and Shorter, Aaron and Shutty, Noah and Shvarts, Vladimir and Singh, Shraddha and Sivak, Volodymyr and Skruzny, Jindra and Small, Spencer and Smelyanskiy, Vadim and Smith, W. Clarke and Somma, Rolando D. and Springer, Sofia and Sterling, George and Strain, Doug and Suchard, Jordan and Szasz, Aaron and Sztein, Alex and Thor, Douglas and Torres, Alfredo and Torunbalci, M. Mert and Vaishnav, Abeer and Vargas, Justin and Vdovichev, Sergey and Vidal, Guifre and Villalonga, Benjamin and Heidweiller, Catherine Vollgraff and Waltman, Steven and Wang, Shannon X. and Ware, Brayden and Weber, Kate and Weidel, Travis and White, Theodore and Wong, Kristi and Woo, Bryan W. K. and Xing, Cheng and Yao, Z. Jamie and Yeh, Ping and Ying, Bicheng and Yoo, Juhwan and Yosri, Noureldin and Young, Grayson and Zalcman, Adam and Zhang, Yaxing and Zhu, Ningfeng and Zobrist, Nicholas},
   year={2024},
   month=dec, pages={920–926} }

@article{kim2023scalable,
  title={Scalable error mitigation for noisy quantum circuits produces competitive expectation values},
  author={Kim, Youngseok and Wood, Christopher J and Yoder, Theodore J and Merkel, Seth T and Gambetta, Jay M and Temme, Kristan and Kandala, Abhinav},
  journal={Nature Physics},
  volume={19},
  number={5},
  pages={752--759},
  year={2023},
  publisher={Nature Publishing Group UK London}
}

@article{piveteau2021error,
  title={Error mitigation for universal gates on encoded qubits},
  author={Piveteau, Christophe and Sutter, David and Bravyi, Sergey and Gambetta, Jay M and Temme, Kristan},
  journal={Physical review letters},
  volume={127},
  number={20},
  pages={200505},
  year={2021},
  publisher={APS}
}

@article{Takagi_2022,
   title={Fundamental limits of quantum error mitigation},
   volume={8},
   ISSN={2056-6387},
   url={http://dx.doi.org/10.1038/s41534-022-00618-z},
   DOI={10.1038/s41534-022-00618-z},
   number={1},
   journal={npj Quantum Information},
   publisher={Springer Science and Business Media LLC},
   author={Takagi, Ryuji and Endo, Suguru and Minagawa, Shintaro and Gu, Mile},
   year={2022},
   month=sep }

@article{Huggins_2021,
   title={Virtual Distillation for Quantum Error Mitigation},
   volume={11},
   ISSN={2160-3308},
   url={http://dx.doi.org/10.1103/PhysRevX.11.041036},
   DOI={10.1103/physrevx.11.041036},
   number={4},
   journal={Physical Review X},
   publisher={American Physical Society (APS)},
   author={Huggins, William J. and McArdle, Sam and O’Brien, Thomas E. and Lee, Joonho and Rubin, Nicholas C. and Boixo, Sergio and Whaley, K. Birgitta and Babbush, Ryan and McClean, Jarrod R.},
   year={2021},
   month=nov }

@article{van_den_Berg_2023,
   title={Probabilistic error cancellation with sparse Pauli–Lindblad models on noisy quantum processors},
   volume={19},
   ISSN={1745-2481},
   url={http://dx.doi.org/10.1038/s41567-023-02042-2},
   DOI={10.1038/s41567-023-02042-2},
   number={8},
   journal={Nature Physics},
   publisher={Springer Science and Business Media LLC},
   author={van den Berg, Ewout and Minev, Zlatko K. and Kandala, Abhinav and Temme, Kristan},
   year={2023},
   month=may, pages={1116–1121} }

@article{Temme_2017,
   title={Error Mitigation for Short-Depth Quantum Circuits},
   volume={119},
   ISSN={1079-7114},
   url={http://dx.doi.org/10.1103/PhysRevLett.119.180509},
   DOI={10.1103/physrevlett.119.180509},
   number={18},
   journal={Physical Review Letters},
   publisher={American Physical Society (APS)},
   author={Temme, Kristan and Bravyi, Sergey and Gambetta, Jay M.},
   year={2017},
   month=nov }

@misc{mohammadipour2025directanalysiszeronoiseextrapolation,
      title={Direct Analysis of Zero-Noise Extrapolation: Polynomial Methods, Error Bounds, and Simultaneous Physical-Algorithmic Error Mitigation}, 
      author={Pegah Mohammadipour and Xiantao Li},
      year={2025},
      eprint={2502.20673},
      archivePrefix={arXiv},
      primaryClass={quant-ph},
      url={https://arxiv.org/abs/2502.20673}, 
}

@article{Campagne_Ibarcq_2020,
   title={Quantum error correction of a qubit encoded in grid states of an oscillator},
   volume={584},
   ISSN={1476-4687},
   url={http://dx.doi.org/10.1038/s41586-020-2603-3},
   DOI={10.1038/s41586-020-2603-3},
   number={7821},
   journal={Nature},
   publisher={Springer Science and Business Media LLC},
   author={Campagne-Ibarcq, P. and Eickbusch, A. and Touzard, S. and Zalys-Geller, E. and Frattini, N. E. and Sivak, V. V. and Reinhold, P. and Puri, S. and Shankar, S. and Schoelkopf, R. J. and Frunzio, L. and Mirrahimi, M. and Devoret, M. H.},
   year={2020},
   month=aug, pages={368–372} }

@article{liu_2025,
  title = {Virtual Channel Purification},
  author = {Liu, Zhenhuan and Zhang, Xingjian and Fei, Yue-Yang and Cai, Zhenyu},
  journal = {PRX Quantum},
  volume = {6},
  issue = {2},
  pages = {020325},
  numpages = {33},
  year = {2025},
  month = {May},
  publisher = {American Physical Society},
  doi = {10.1103/PRXQuantum.6.020325},
  url = {https://link.aps.org/doi/10.1103/PRXQuantum.6.020325}
}

@misc{liu2025quantumerrormitigationsampling,
      title={Quantum Error Mitigation for Sampling Algorithms}, 
      author={Kecheng Liu and Zhenyu Cai},
      year={2025},
      eprint={2502.11285},
      archivePrefix={arXiv},
      primaryClass={quant-ph},
      url={https://arxiv.org/abs/2502.11285}, 
}

@article{Koczor_2021,
  title = {Exponential Error Suppression for Near-Term Quantum Devices},
  author = {Koczor, B\'alint},
  journal = {Phys. Rev. X},
  volume = {11},
  issue = {3},
  pages = {031057},
  numpages = {30},
  year = {2021},
  month = {Sep},
  publisher = {American Physical Society},
  doi = {10.1103/PhysRevX.11.031057},
  url = {https://link.aps.org/doi/10.1103/PhysRevX.11.031057}
}

@article{Li_2017,
  title = {Efficient Variational Quantum Simulator Incorporating Active Error Minimization},
  author = {Li, Ying and Benjamin, Simon C.},
  journal = {Phys. Rev. X},
  volume = {7},
  issue = {2},
  pages = {021050},
  numpages = {14},
  year = {2017},
  month = {Jun},
  publisher = {American Physical Society},
  doi = {10.1103/PhysRevX.7.021050},
  url = {https://link.aps.org/doi/10.1103/PhysRevX.7.021050}
}

@misc{araki2025correctingquantumerrorsusing,
      title={Correcting quantum errors using a classical code and one additional qubit}, 
      author={Tenzan Araki and Joseph F. Goodwin and Zhenyu Cai},
      year={2025},
      eprint={2510.05008},
      archivePrefix={arXiv},
      primaryClass={quant-ph},
      url={https://arxiv.org/abs/2510.05008}, 
}

@article{Vodola_2022,
   title={Fundamental thresholds of realistic quantum error correction circuits from classical spin models},
   volume={6},
   ISSN={2521-327X},
   url={http://dx.doi.org/10.22331/q-2022-01-05-618},
   DOI={10.22331/q-2022-01-05-618},
   journal={Quantum},
   publisher={Verein zur Forderung des Open Access Publizierens in den Quantenwissenschaften},
   author={Vodola, Davide and Rispler, Manuel and Kim, Seyong and Müller, Markus},
   year={2022},
   month=jan, pages={618} }

@article{Dennis_2002,
   title={Topological quantum memory},
   volume={43},
   ISSN={1089-7658},
   url={http://dx.doi.org/10.1063/1.1499754},
   DOI={10.1063/1.1499754},
   number={9},
   journal={Journal of Mathematical Physics},
   publisher={AIP Publishing},
   author={Dennis, Eric and Kitaev, Alexei and Landahl, Andrew and Preskill, John},
   year={2002},
   month=sep, pages={4452–4505} }

@article{Honecker_2001,
   title={Universality Class of the Nishimori Point in the 2D +/- Random-Bond Ising Model},
   volume={87},
   ISSN={1079-7114},
   url={http://dx.doi.org/10.1103/PhysRevLett.87.047201},
   DOI={10.1103/physrevlett.87.047201},
   number={4},
   journal={Physical Review Letters},
   publisher={American Physical Society (APS)},
   author={Honecker, A. and Picco, M. and Pujol, P.},
   year={2001},
   month=jul }

@inproceedings{Kuo_2022,
   title={Comparison of 2D topological codes and their decoding performances},
   url={http://dx.doi.org/10.1109/ISIT50566.2022.9834489},
   DOI={10.1109/isit50566.2022.9834489},
   booktitle={2022 IEEE International Symposium on Information Theory (ISIT)},
   publisher={IEEE},
   author={Kuo, Kao-Yueh and Lai, Ching-Yi},
   year={2022},
   month=jun, pages={186–191} }

@misc{chen2025fasterprobabilisticerrorcancellation,
      title={Faster Probabilistic Error Cancellation}, 
      author={Yi-Hsiang Chen},
      year={2025},
      eprint={2506.04468},
      archivePrefix={arXiv},
      primaryClass={quant-ph},
      url={https://arxiv.org/abs/2506.04468}, 
}

@unpublished{Dai2025StratifiedQPD,
  author = {Dai, Joshua and others},
  title  = {Stratified Sampling for Classical Randomisation Schemes},
  note   = {Manuscript in preparation; to appear on arXiv},
  year   = {2026}
}

@article{caiPracticalFrameworkQuantum2021,
  title = {A Practical Framework for Quantum Error Mitigation},
  author = {Cai, Zhenyu},
  year = 2021,
  month = oct,
  number = {arXiv:2110.05389 [quant-ph]},
  primaryclass = {quant-ph},
  publisher = {arXiv},
  doi = {10.48550/arXiv.2110.05389},
  url = {http://arxiv.org/abs/2110.05389},
  journal = {arXiv}
}

@article{caiMultiexponentialErrorExtrapolation2021,
  ids = {caiMultiexponentialErrorExtrapolation2020},
  title = {Multi-Exponential Error Extrapolation and Combining Error Mitigation Techniques for {{NISQ}} Applications},
  author = {Cai, Zhenyu},
  year = 2021,
  month = may,
  journal = {npj Quantum Information},
  volume = {7},
  pages = {80},
  publisher = {Nature Publishing Group},
  issn = {2056-6387},
  doi = {10.1038/s41534-021-00404-3},
  url = {https://www.nature.com/articles/s41534-021-00404-3}
}

@article{bonet-monroigLowcostErrorMitigation2018,
  title = {Low-Cost Error Mitigation by Symmetry Verification},
  author = {{Bonet-Monroig}, X. and Sagastizabal, R. and Singh, M. and O'Brien, T. E.},
  year = 2018,
  month = dec,
  journal = {Physical Review A},
  volume = {98},
  number = {6},
  pages = {62339},
  doi = {10.1103/PhysRevA.98.062339},
  url = {https://link.aps.org/doi/10.1103/PhysRevA.98.062339}
}

@article{mcardleErrormitigatedDigitalQuantum2019,
  title = {Error-Mitigated Digital Quantum Simulation},
  author = {McArdle, Sam and Yuan, Xiao and Benjamin, Simon},
  year = 2019,
  month = may,
  journal = {Physical Review Letters},
  volume = {122},
  number = {18},
  pages = {180501},
  doi = {10.1103/PhysRevLett.122.180501},
  url = {https://link.aps.org/doi/10.1103/PhysRevLett.122.180501}
}

@article{zhuReversingUnknownQuantum2024,
  title = {Reversing Unknown Quantum Processes via Virtual Combs for Channels with Limited Information},
  author = {Zhu, Chengkai and Mo, Yin and Chen, Yu-Ao and Wang, Xin},
  year = 2024,
  month = jul,
  journal = {Physical Review Letters},
  volume = {133},
  number = {3},
  pages = {30801},
  issn = {0031-9007, 1079-7114},
  doi = {10.1103/PhysRevLett.133.030801},
  url = {https://link.aps.org/doi/10.1103/PhysRevLett.133.030801}
}

@article{oxford_arc, 
    publisher={Zenodo}, 
    url = {https://doi.org/10.5281/zenodo.22558},
    journal ={University of Oxford Advanced Research Computing},
    author={Richards, Andrew}, 
    year={2015}, 
    month={Aug}
}
\end{document}